\documentstyle[epsfig]{lamuphys}
\makeatletter
\let\chapter\hid@chapter
\makeatother

\catcode`\@=11
\newcommand{\mybibitem}[2]{\bibitem{[#2]}{#1}{[#2]}}
\newcommand{\vectorp}{\mbox{\boldmath{$p$}}}
\newcommand{\vectorr}{\mbox{\boldmath{$r$}}}

\begin{document}
\setcounter{page}{0}
  
\titlerunning{The Solar Neutrino Problem}
\title{The Solar Neutrino Problem and Solar Neutrino 
Oscillations in Vacuum and in Matter} 

\author{S.T. Petcov}
\institute{Scuola Internazionale Superiore di Studi Avanzati, and \\
Istituto Nazionale di Fisica Nucleare, 
Sezione di Trieste, I--34013 Trieste, Italy;\\
Also at: Institute of 
Nuclear Research and Nuclear Energy, Bulgarian Academy of Sciences, 
1784 Sofia, Bulgaria.}
\maketitle
\begin{abstract}
The solar neutrino problem is reviewed and the possible vacuum 
oscillation and MSW solutions of the problem are considered.
\end{abstract}

\setcounter{footnote}{0}
\renewcommand{\thefootnote}{\alph{footnote}}

\newcommand{\be}[1]{\begin{equation} \label{(#1)}}
\newcommand{\ee}{\end{equation}}
\newcommand{\ba}[1]{\begin{eqnarray} \label{(#1)}}
\newcommand{\ea}{\end{eqnarray}}
\newcommand{\nn}{\nonumber}
\newcommand{\rf}[1]{(\ref{(#1)})}

\section{Introduction}

The problem of neutrino mass is the central problem of present 
day neutrino physics and one of the central problems of contemporary elementary
particle physics.\footnote{The neutrino mass problem, 
the phenomenological implications of the nonzero neutrino mass and 
lepton mixing hypothesis, the properties of massive Dirac and massive 
Majorana neutrinos, the neutrino mass generation 
in the contemporary gauge theories 
of the electroweak interaction as well the role massive neutrinos
can play in astrophysics and cosmology are the subject of a number
of review articles and books: see, e.g., \cite{NMP1}--\cite{NMP6}.} 
The existence of nonzero neutrino masses is typically correlated
in the modern theories of the elementary particle interactions with
nonconservation of the additive lepton charges, $L_e,~L_{\mu}$ and $L_{\tau}$
(see, e.g., \cite{NMP1}).
The rather stringent experimental limits on neutrino masses obtained so far
together with cosmological arguments imply (see, e.g., \cite{JEnu96}) 
that if nonzero, the masses of the flavour neutrinos must be
by many orders of magnitude smaller 
than the masses of the corresponding charged lepton and quarks
belonging to the same family as the neutrino. 
The extraordinary smallness
of the neutrino masses is related in the modern theories of electroweak
interactions with massive neutrinos 
to the existence of new mass scales in these theories.
Thus, the studies of the neutrino mass problem 
are intimately related to the 
studies of the basic symmetries
of electroweak interactions; they are also closely connected 
with the investigations of the possibility of existence 
of new scales in elementary particle physics. 
Correspondingly, the experiments searching for effects
of nonzero neutrino masses and lepton mixing are 
actually testing the fundamental symmetries 
of the electroweak interactions. These experiments
are also searching for indirect evidences  
for existence of new scales in physics. 

Neutrinos are massless particles in the
standard (Glashow-Salam-Weinberg) theory 
(ST) of the electroweak interactions. The observation
of effects of nonzero neutrino masses and lepton mixing would be a very
strong indication for existence of new physics beyond that predicted by the 
standard theory. The studies of the neutrino mass problem can lead to a progress 
in our understanding of the nature of the dark matter in the Universe as well
\cite{JPrimacknu96}. 

One of the most interesting and beautiful phenomenological consequences of 
the nonzero neutrino mass and lepton mixing hypothesis are the oscillations of
neutrinos \cite{Pont57}, i.e., 
transitions in flight between different types of neutrinos,
$\nu_{e} \leftrightarrow \nu_{\mu}$ and/or 
$\nu_{e} \leftrightarrow \nu_{\tau}$ and/or
$\nu_{\mu} \leftrightarrow \nu_{\tau}$, and antineutrinos, 
$\bar{\nu}_{e} \leftrightarrow \bar{\nu}_{\mu}$ and/or
$\bar{\nu}_{e} \leftrightarrow \bar{\nu}_{\tau}$ and/or
$\bar{\nu}_{\mu} \leftrightarrow \bar{\nu}_{\tau}$.
If, for example, a beam of $\nu_e$ neutrinos is produced by some source,
at certain distance R from the source the beam will acquire a substantial
$\nu_{\mu}$ component if $\nu_{e} \leftrightarrow \nu_{\mu}$ oscillations 
take place. The probability to find $\nu_{\mu}$ at distance R 
from the source of $\nu_e$ when the neutrinos propagate in vacuum and the massive
neutrinos are relativistic, 
$P(\nu_{e} \rightarrow \nu_{\mu})$, is a
function of the neutrino energy E, the differences of the squares of the 
masses $m_k$ of the neutrinos $\nu_k$ 
having definite mass in vacuum, $\Delta m^2_{jk} = m^2_{j} - m^2_{k}$,
and of the elements of the lepton mixing matrix $U$ (see, e.g., \cite{NMP1,NMP2}). 
  
At present we have several indications that neutrinos 
indeed take part in oscillations, which suggest that 
neutrinos have nonzero 
masses and that lepton mixing exists.
One of the indications comes from the 
results of the LSND neutrino oscillation experiment
performed at the Los Alamos meson factory \cite{LSND}. 
The events observed in this experiment
can be interpreted as being due to $\bar{\nu}_{\mu} \leftrightarrow \bar{\nu}_{e}$
oscillations with $\Delta m^2 \sim~\mbox{few}~eV^2$ and 
$\sin^22\theta \sim~\mbox{few}\times 10^{-3}$, where 
$\Delta m^2$ and $\sin^22\theta$ are the two parameters -- the neutrino 
mass squared difference and the neutrino mixing angle, 
which characterize the oscillations in the simplest case.
The second indication is usually referred to as the atmospheric neutrino
problem or anomaly \cite{ATNUP1,ATNUP2}: 
the ratio of the $\mu-$like and $e-$like events produced
respectively by the fluxes of $(\nu_{\mu} + \bar{\nu}_{\mu})$ and   
$(\nu_{e} + \bar{\nu}_{e})$ atmospheric neutrinos with energies
$\sim (0.2 - 10.0)~$GeV, detected in Kamiokande, IMB, Soudan and Super-Kamiokande
experiments, is smaller than the theoretically predicted ratio.
The atmospheric neutrino data can be explained by 
$\nu_{\mu} \leftrightarrow \nu_{\tau}$ and
$\bar{\nu}_{\mu} \leftrightarrow \bar{\nu}_{\tau}$ oscillations with
$\Delta m^2 \sim (10^{-3} - 10^{-2})~eV^2$ and a relatively large value of 
$\sin^22\theta$ close to 1. 

The amount of the solar neutrino data available at present, 
the numerous nontrivial checks of the
functioning of the solar neutrino detectors that have been 
and are being performed,
together with recent results in the field of solar modeling
associated, in particular, with the publication of new more precise
helioseismological data and their interpretation, suggest, however, that 
the most substantial evidence for existence of nonzero 
neutrino masses and lepton mixing
comes at present from the results of the solar neutrino experiments.
In view of this
we will devote the present lectures
to the solar neutrino problem and its possible neutrino oscillation
solutions. 

The ``story'' of solar neutrinos begins, to our knowledge, in 1946 with the
well-known article by B. Pontecorvo \cite{Pont46}, published only as a report of the 
Chalk River Laboratory (in Canada). In it Pontecorvo suggested that reactors 
and the Sun are copious sources of neutrinos. On the basis of neutrino flux 
and interaction cross-section estimates he concluded \cite{Pont46} that the 
experimental detection of neutrinos emitted by a reactor (i.e., the 
observation of a reaction caused by neutrinos) is feasible, while the detection 
of solar neutrinos can be very difficult (but not impossible). In the same 
article the radiochemical method of detection of neutrinos was proposed. As 
a possible concrete realization of the method, a detector based on the Cl--Ar 
reaction ${\rm \nu_{e} +~^{37}Cl \rightarrow~^{37}Ar + e^{-}}$ was discussed.
The possibility to use the Cl--Ar method for detection of neutrinos was 
further studied in 1949 by Alvarez \cite{Alv49}. A Cl--Ar detector for observation of 
solar neutrinos was eventually built
by Davis and his collaborators \cite{Davis68}. The epic Homestake experiment of 
Davis and collaborators, in which for the first time neutrinos emitted by the 
Sun were detected, began to operate in 1967 and still continues to provide 
data. It was realized in 1967 as well \cite{Pont67}
that the measurements of the 
solar neutrino flux can give unique information
not only about the physical conditions and the nuclear reactions taking place
in the central part of the Sun, but also about the neutrino intrinsic 
properties.

The solar neutrino problem emerged in the 70'ies
as a discrepancy between the results of the 
Davis et al. experiment \cite{Davis68,Davis} and the
theoretical predictions for the signal in this experiment \cite{BU82}, based on
detailed solar model calculations. The hypothesis of unconventional behaviour
of the solar $\nu_e$ on their way to the Earth (as like, e.g., vacuum 
oscillations \cite{Pont57,Pont67} $\nu_e \leftrightarrow \nu_{\mu (\tau)}$ and/or 
$\nu_e \leftrightarrow \nu_s$, $\nu_s$ being a sterile neutrino, 
etc.) provided a natural explanation of 
the deficiency of solar neutrinos reported by
Davis et al. However, as the fraction of the solar $\nu_e$ flux to which the
experiment of Davis et al. is sensitive (neutrinos with energy E $\geq$ 0.814 
MeV) was known \cite{BU82} i) to be produced in a chain of nuclear 
reactions (representing a branch of the pp cycle) which play a
minor role in the physics of the Sun and whose cross--sections cannot  
all be measured directly in the relevant energy range on Earth, and ii) to be
extremely sensitive to the predicted value of the central temperature, T$_{c}$,
in the Sun (scaling as T$^{24}_{c}$), the possibility of an alternative
(astrophysics, nuclear physics) explanation of the Davis et al. results could
not be excluded.

In 1986 an independent measurement of the high energy part (E $\geq$ 7.5
MeV) of the flux of solar neutrinos was successfully undertaken by the Kamiokande
II collaboration using a completely different experimental technique; in
1990 the measurements were continued by the
Kamiokande III group with an improved version of the Kamiokande II detector
\cite{Kam}. At the beginning of the 90'ies two new experiments, 
SAGE \cite{SAGE} and GALLEX
\cite{GALLEX}, sensitive to the low energy part (E $\geq$ 0.233 MeV) of the solar 
neutrino flux, began to operate and to provide qualitatively new data. 
The Kamiokande III detector was succeeded by an approximately 30 times bigger
version called appropriately ``Super-Kamiokande'', which 
began solar and atmospheric neutrino detection on April 1, 1996 \cite{SK}.
The data obtained since 1986 did not alleviate the solar neutrino problem -- 
on the contrary, they made the case for existence of solar neutrino deficit
even stronger.

At the same time
considerable efforts were also made to understand better the potential sources
and the possible magnitude of the uncertainties in the theoretical predictions
for the signals in the indicated solar neutrino detectors, and to develop
improved, physically more precise solar models on the basis of which the
predictions are obtained \cite{JNB}. 
Remarkable progress in this direction was
made in the last several years with the development of the solar models which 
include the diffusion of helium and the heavy elements in the Sun 
\cite{BP92}--\cite{DS96},  
as well as with the appearance of
new more precise helioseismological data permitting new 
critical tests of the solar models to be performed
\cite{HELIOSEIS0}--\cite{HELIOSEIS3}. 

With the accumulation of more data and the
developments in the theory certain aspects of the solar neutrino problem
changed and new aspects appeared. 

In the present lectures we shall review the current
status of the solar neutrino problem. We shall also review the
status of the neutrino physics solutions of the problem based on the hypotheses
of vacuum oscillations \cite{Pont57} or of matter-enhanced transitions \cite{MSW1,MSW2} 
of solar neutrinos. 

\section{The Data and the Solar Model Predictions}

We begin with a brief summary of relevant solar model predictions and of
the solar neutrino data. According to the existing models of the Sun 
\cite{JNB},
the solar $\nu_e$ flux consists of several components, six of which are
relevant to our discussion: 
\begin{enumerate}
\item[i)] the least energetic pp neutrinos (E $\leq$ 0.420
MeV, average energy $\bar{{\rm E}} = 0.265~$MeV),
\item[ii)] the intermediate energy
monoenergetic $^{7}$Be neutrinos (E$=$0.862 MeV (89.7\% of the flux), 
0.384 MeV 
(10.3\% of the flux)),
\item[iii)]  the higher energy $^{8}$B neutrinos (E $\leq$ 14.40
MeV, $\bar{{\rm E}} = 6.71~$MeV), and three additional intermediate energy
components, namely, 
\item[iv)] the monoenergetic pep neutrinos (E$=$1.442 MeV), and 
the continuous spectrum CNO neutrinos produced in the $\beta^{+}-$decays 
\item[v)] of $^{13}$N (E $\leq$ 1.199 MeV, $\bar{{\rm E}} = 0.707~$MeV), and  
\item[vi)] of $^{15}$O (E $\leq$ 1.732 MeV, $\bar{{\rm E}} = 0.997~$MeV). 
\end{enumerate}
\begin{figure}[tbp]
\begin{center}
\epsfig{file=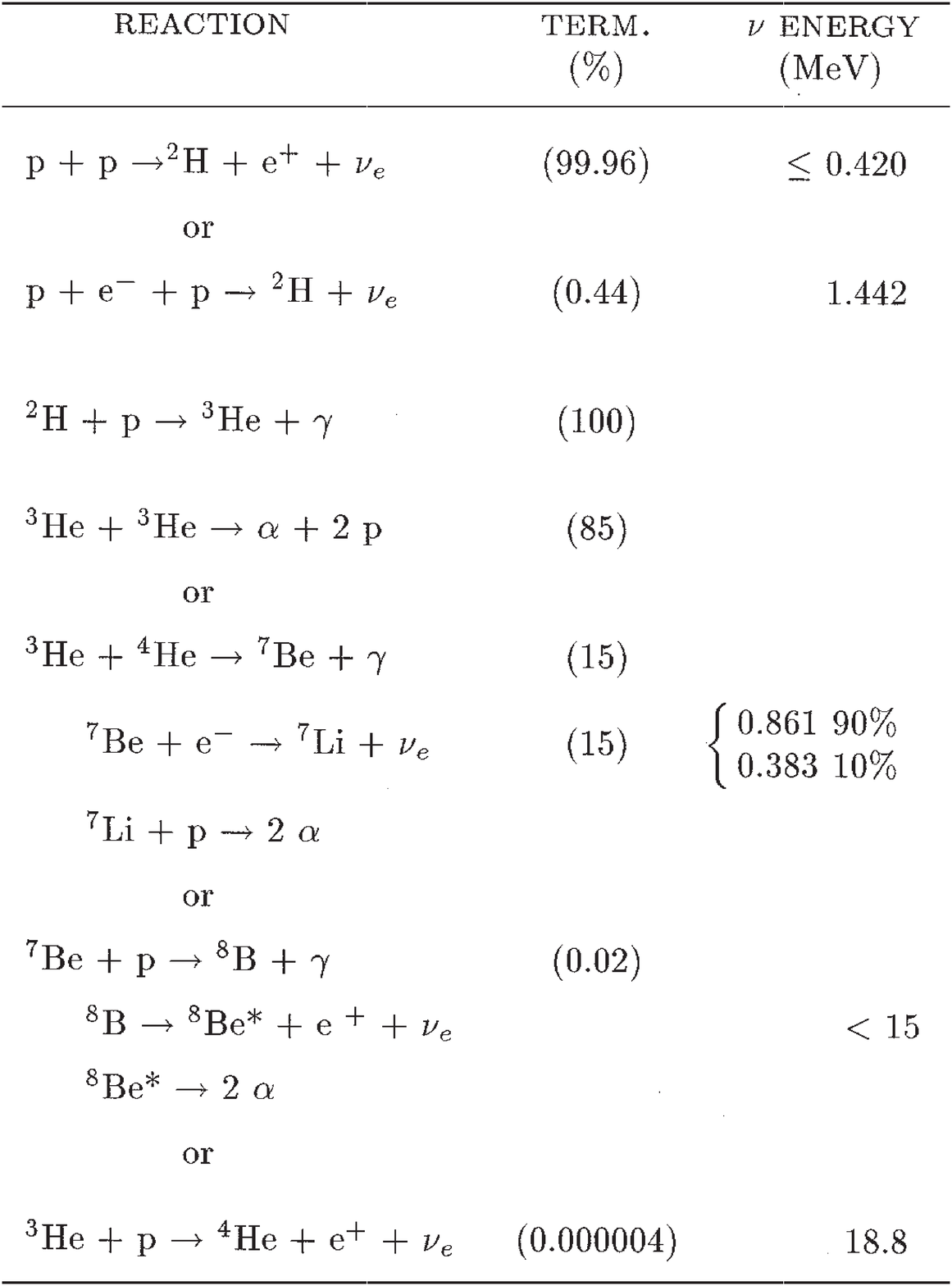,width=10cm}
\end{center}
\caption{The nuclear reactions of the pp--chain in the Sun.}
\end{figure}
The pp, pep, $^{7}$Be and $^{8}$B
neutrinos are produced in a set of nuclear reactions shown in Fig. 1.
These make part of three major cycles (the pp-cycles) of nuclear fusion reactions
in which effectively 4 protons burn into $^{4}$He with emission of
two positrons and two neutrinos, generating approximately 98\% of the solar energy:
$$4{\rm p} \rightarrow~^{4}{\rm He} + 2{\rm e}^{+} + 2\nu_e.~~\eqno(1)$$
The first (pp-I) cycle (or chain) 
begins with the p-p (or p-e$^{-}$-p) fusion into
deuterium and ends with the reaction $^{3}{\rm He} + ^{3}{\rm He} 
\rightarrow~ ^{4}{\rm He} + 2p$. The second (pp-II) and the third (pp-III) cycles 
begin with the production of $^{7}$Be in 
$^{3}{\rm He} + ^{4}{\rm He}$ fusion and end
respectively with the processes $^{7}{\rm Li} + p \rightarrow~2 ^{4}{\rm He}$ 
and $^{8}{\rm B} \rightarrow~2 ^{4}{\rm He} + {\rm e}^{+} + \nu_e$ (see Fig. 1).

The CNO neutrinos are produced in the CNO-cycle of reactions, which,
according to the present day understanding, 
plays minor role in the energetics of the Sun: 
$^{12}{\rm C} + p \rightarrow~ ^{13}{\rm N} + \gamma$,
$^{13}{\rm N} \rightarrow~ ^{13}{\rm C} + {\rm e}^{+} + \nu_e$,
$^{13}{\rm C} + p \rightarrow~ ^{14}{\rm N} + \gamma$,
$^{14}{\rm N} + p \rightarrow~ ^{15}{\rm O} + \gamma$,
$^{15}{\rm O} \rightarrow~ ^{15}{\rm N} + {\rm e}^{+} + \nu_e$ and
$^{15}{\rm N} + p \rightarrow~ ^{12}{\rm C} + ^{4}{\rm He}$.
The pp, pep, $^{7}$Be and $^{8}$B neutrino spectra are depicted in Fig. 2.
Let us note that the shapes of the continuous spectra of the pp, $^{8}$B and the 
CNO neutrinos are to a high degree of accuracy  
solar physics independent. The total fluxes 
of the pp, pep, $^{7}$Be, $^{8}$B and CNO 
neutrinos depend, although to a different degree, on the physical 
conditions in the central part of the Sun \cite{JNB} (see further).  

\begin{figure}[htbp]
\begin{center}
\epsfig{file=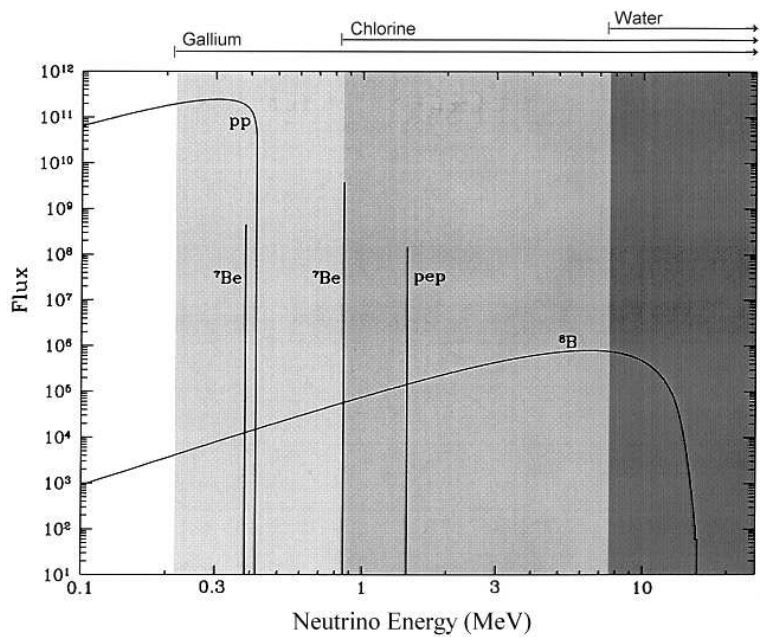,width=10cm}
\end{center}
\caption{The spectra of the pp, pep, $^7$Be and $^8$B neutrino fluxes 
(from [23]
). Shown are also the ranges of solar neutrino
energies to which the Ga - Ge, Cl - Ar and Kamiokande II and III 
experiments are sensitive.}
\end{figure}

Solar neutrinos are produced in the central solar region 
(which practically coincides with energy production region) with radius
$r_{\nu} \cong 0.25~R_{\odot}$, $R_{\odot} = 6.96\times 10^{5}~{\rm km}$ being 
the radius of the Sun. The dependence of the 
source-strength functions for the  
pp, $^{7}$Be and $^{8}$B neutrinos 
on the distance from the center of the Sun, $r$,  
is shown in Fig. 3. As this figure illustrates, 
the major part of the
$^{8}$B neutrinos flux is generated in a rather small region, 
$r \la 0.10~R_{\odot}$ close to the center of the Sun; the 
region of production of $^{7}$Be neutrinos extends to 
$r \cong 0.15~R_{\odot}$, while the region of the pp neutrino production is
the largest extending to $r \cong 0.25~R_{\odot}$. 

\begin{figure}[htb]
\begin{center}
\epsfig{file=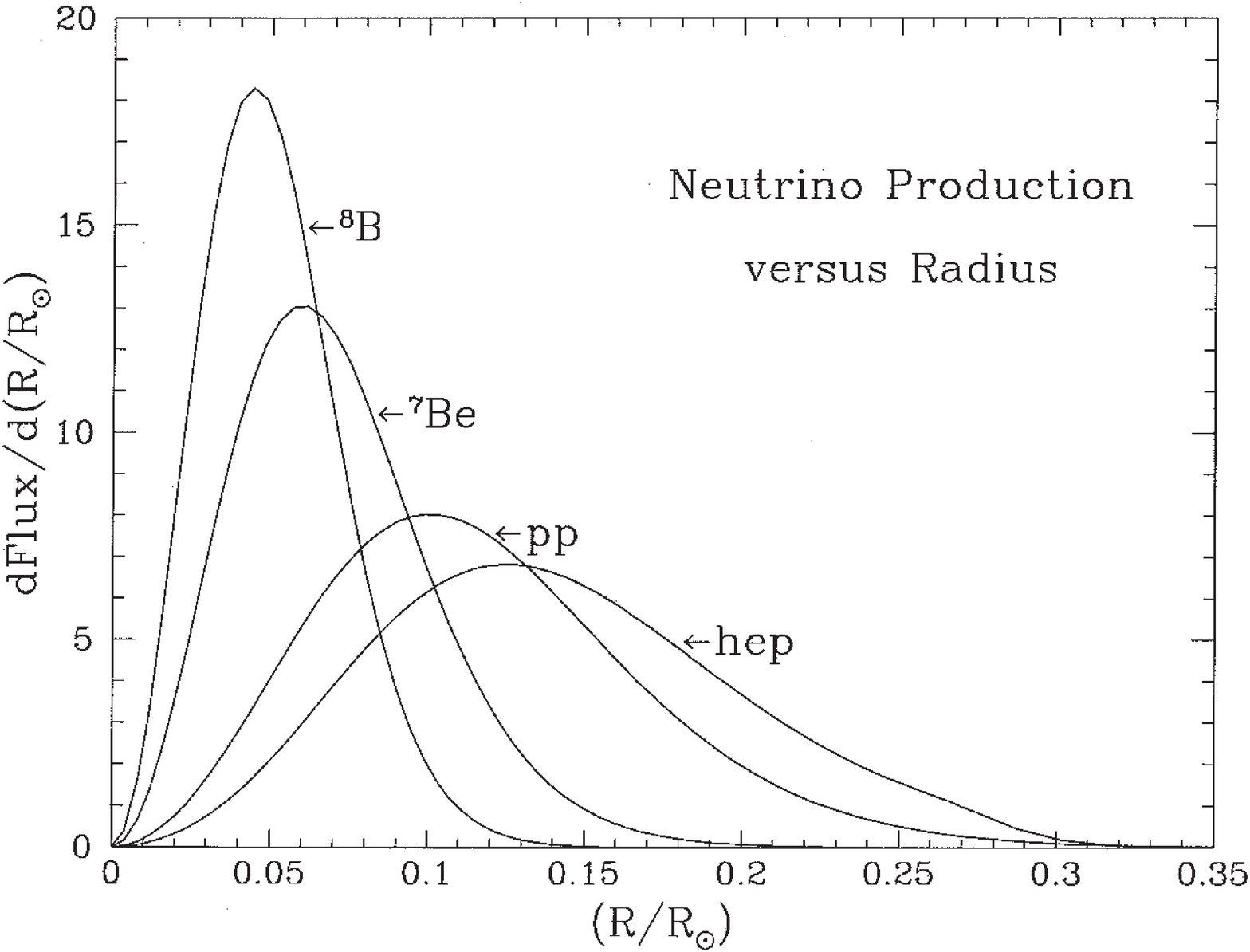,width=9cm}
\end{center}
\caption{The production of the fluxes of pp, $^{7}$Be and $^{8}$B
neutrinos as a function of the distance R from the center of the Sun 
(the latter is expressed in
units of the solar radius ${\rm R_{\odot}}$) (from [23]
).}
\end{figure}

Three different methods of solar neutrino detection have been and are being
used in the six solar neutrino experiments 
\cite{Davis}, \cite{Kam}--\cite{SK}
that have provided data so far: the 
Cl--Ar method proposed by Pontecorvo 
\cite{Pont46} -- in the experiment of Davis et al. \cite{Davis68,Davis}, 
the $\nu - e^{-}$ elastic 
scattering reaction -- in the (Super{-}) Kamiokande 
experiments \cite{Kam,SK}, and the 
radiochemical Ga--Ge method -- in 
SAGE \cite{SAGE} and GALLEX \cite{GALLEX} experiments. 

The threshold energy of the reaction 
$\nu_e + ^{37}{\rm Cl} \rightarrow e^{-} + ^{37}{\rm Ar}$ on which the Cl--Ar
method is based, is ${\rm E_{th}(Cl) = 0.814~MeV}$. Consequently, the pp 
neutrinos do not give contributions in the signal in 
the Davis et al. detector. Inspecting the
predictions of all solar models presently discussed in the literature one finds
that the major contribution to the signal in the Cl--Ar experiment, between
64\% and 79\%, should be due to the $^{8}$B neutrinos; the $^{7}$Be neutrinos
are predicted to generate between 22\% and 13\% of the total signal, and the
pep and the CNO neutrinos -- between 13\% and 8\%. With a threshold neutrino
energy, E$_{{\rm th}}({\rm K})$, first of 9.5 MeV (7.5 MeV) and subsequently  
reduced to 7.5 MeV and further to 7.3 MeV (6.5 MeV), the (Super{-}) 
Kamiokande experiments can 
detect only the higher energy $^{8}$B
component of the solar $\nu_e$ flux. 

Having the lowest threshold energy
E$_{{\rm th}}({\rm Ga}) =~ $0.233 MeV, the Ga--Ge detectors GALLEX and SAGE are
sensitive to all six components of the flux considered above. Moreover, the
major part of the signal in these detectors, between 51\% and 63\%, is
predicted to be produced by the pp neutrinos; the $^{7}$Be neutrinos are
expected to generate between 28\% and 24\% of the total signal, the 
$^{8}$B neutrinos -- between 12\% and 5\%, and the CNO neutrinos -- around 
(5--7)\%. 

The above analysis implies that the Cl--Ar and Kamiokande 
experiments on one side, and the Ga--Ge experiments on the other, are most 
sensitive to very
different components of the solar neutrino flux: the former -- to the $^{8}$B
neutrinos, and the latter -- to the pp neutrinos. The $^{7}$Be neutrinos are
predicted to give the second largest (and non-negligible) contributions to the
signals in both Cl--Ar and Ga--Ge experiments.

Let us turn next to the data. The average rate of $^{37}$Ar production by
solar neutrinos, $\bar{\rm R}$(Ar), observed in the experiment of Davis et al. 
in the period 1971--1996 (altogether $\sim 800$ solar $\nu_e$ induced events
registered) is \cite{Davis} 
\begin{equation}
(2.56 \pm 0.16 \pm 0.14)\hskip 0.2cm {\rm SNU}.
\end{equation}           
Here (and in the experimental results we quote further) the 
first error is statistical (1 s. d.) and the second error is systematic. 
The flux of $^{8}$B neutrinos, $\Phi_{{\rm B}}$, 
measured by the Kamiokande experiments reads \cite{Kam}
\begin{equation}
\bar\Phi^{K}_{{\rm B}} = 
(2.80 \pm 0.19 \pm 0.33) \times 10^{6}~{\rm cm}^{-2}{\rm sec}^{-1}.
\end{equation}
The result is based on a statistics of about 600 events accumulated
by the two experiments in 2079 days of measurements in the period 1986 -- 1996.

The GALLEX and SAGE experiments began to collect data in 1991 and 1990,
respectively. 
The GALLEX group has registered 
(in 65 runs) approximately 300, and
the SAGE group has registered (in 33 runs)
about 100 solar neutrino induced events. The average rates of $^{71}$Ge 
production by solar neutrinos, $\bar {\rm R}({\rm Ge})$, measured by the two 
collaborations are \cite{GALLEX,SAGE}
\begin{equation}
\bar {\rm R}_{\rm GALLEX}({\rm Ge}) = 
 (76.2~\pm 6.5~\pm 5)\hskip 0.2cm {\rm SNU},
\end{equation}
\begin{equation}
\bar {\rm R}_{{\rm SAGE}}({\rm Ge}) = 
(73~\pm~ 8.5~ ^{+5.2}_{-6.9})\hskip 0.2cm {\rm SNU}.
\end{equation}
Obviously, the results of the two experiments are compatible.
Adding the statistical and the systematical errors in (3) and in  (4) 
in quadratures and combining the two results 
(i.e., taking the weighted average) we find
\begin{equation}
\bar {\rm R}_{{\rm exp}}({\rm Ge}) = (75.4~\pm~7)\hskip 0.2cm {\rm SNU}.
\end{equation}

Recently, the Super-Kamiokande collaboration announced 
results on solar neutrinos based on data collected during a period of
306.3 days. About 4000 (!) solar neutrino induced events have been detected. 
The following value of the $^{8}$B neutrino flux was measured \cite{SK}:  
\begin{equation}
\bar\Phi^{SK}_{{\rm B}} = (2.44~\pm~0.06~^{+0.25}_{-0.09}) 
\times 10^{6}~{\rm cm}^{-2}{\rm sec}^{-1}.
\end{equation}
 
The GALLEX collaboration has successfully performed in 1994 and in 1997 
very important
(and rather spectacular) calibration experiments with an artificially prepared
powerful $^{51}$Cr source of monoenergetic $\nu_e$ (four lines: E = 746 keV
(81\%), 751 keV (9\%), 426 keV (9\%) and 431 keV (1\%)) of known intensity
\cite{GALLEX}. At the beginning of the two exposures the signal due to the 
$^{51}$Cr neutrinos was approximately 15 times bigger than the signal due to
the solar neutrinos. The results of the experiments showed, in particular, that
the efficiency of extraction of the $^{71}$Ge, produced by neutrinos, from the
tank of the detector coincides (within the 10\% error) with the calculated 
one. 

Similar calibration experiment has been successfully completed in 1996 also by 
the SAGE collaboration. These experiments demonstrated, in particular, that the 
Ga-Ge detectors are capable of detecting 
the intermediate energy $^{7}$Be neutrinos with 
a high efficiency. They represent a solid proof that the data on the solar 
neutrinos provided by the two detectors are correct. They also represent 
the first real proof of 
the feasibility of the radiochemical method invented by Pontecorvo \cite{Pont46} for 
detection and quantitative study of solar neutrinos. 

An extensive program of calibration studies of the Super-Kamiokande detector 
is presently being completed. The aim is to reach an accuracy of 1\% in the
measurement of the energy of the recoil e$^{-}$ from the solar neutrino
induced reaction 
$\nu + e^{-} \rightarrow \nu + e^{-}$.

\section{The Data Versus the Solar Model Predictions}
\subsection{Modeling the Sun}

The results of the solar neutrino experiments have to be compared with the
corresponding theoretical predictions. 
Many authors have worked (and many continue to work) in the field of solar 
modeling and have produced predictions
for the values of the pp, $^{7}$Be, $^{8}$B, pep and CNO neutrino fluxes, and
for the signals in the solar neutrino detectors: a rather detailed review of
the results obtained by different authors prior 1992 and the corresponding
references can be found in \cite{BP92}.
The articles 
\cite{TCL}, \cite{Bert}--\cite{DS96} 
describe
some of the models proposed after 1992. 
Most persistently solar models with
increasing sophistication and precision, aiming to account for and/or 
reproduce with sufficient accuracy the physical conditions and the possible processes
taking place in the inner parts of the Sun have been developed starting from
1964 by John Bahcall and his collaborators \cite{JNB}.

The solar models are based on the standard assumptions of hydrostatic
equilibrium and energy conservation made in the theory
of stellar evolution. Several additional ingredients are needed to determine
the physical structure of the Sun and its evolution in time \cite{JNB,Cast2}:
\begin{enumerate}
\item the initial chemical composition, 
\item the equation of state,
\item the rate of energy production per unit mass as a function of the density $\rho$,
temperature T and chemical composition $\mu$, 
\item the radiative opacity $\kappa$ as a function of the same three quantities, and
\item the mechanism of energy transport.
\end{enumerate}

The initial chemical composition of the Sun is, of course, unknown.
However, the relative abundances of the heavy elements in the initial Sun, 
with the exception of the noble gases and  
C, N, and O,  
are expected to be approximately equal to those found 
in the type I carbonaceous chondrite meteorites 
(see, e.g., \cite{JNB}). 
Using the meteoritic abundances and the 
measured abundances in the solar photosphere which includes
hydrogen, and taking into account the possible change of the heavy
element abundances during the evolution of the Sun, 
permits to fix the present day 
ratio of the heavy element (Z) and 
hydrogen (X) mass fractions, Z/X. The knowledge of Z/X, 
the normalization condition
$X + Y + Z = 1$, where $Y$ is the helium mass fraction, and the requirement
that the solar model reproduces correctly the measured value of
the solar luminosity (see further)
allows to determine the absolute values of the initial solar element abundances
(for further details see \cite{JNB}).
 
The equation of state requires the knowledge of the degree of ionization and
the population of the excited states of all elements present in the Sun. 
Stellar plasma effects which introduce deviations from the perfect 
gas law have to be taken into account as well.
As we have mentioned earlier, the energy is generated in the Sun in
four cycles (or chains) of nuclear fusion reactions in which effectively
4 protons burn into $^{4}$He: 
$4{\rm p} \rightarrow~^{4}{\rm He} + 2{\rm e}^{+} + 2\nu_e$. 
Collective plasma and screening effects have to be accounted for 
in the calculation of the corresponding 
nuclear reaction cross-sections.

The radiative opacity $\kappa$ is determined by the photon mean free path.
It controls the temperature gradient (and therefore the energy flow)
in the radiative zone. The calculation of $\kappa$ requires a detailed
knowledge of all atomic levels in the solar interior as well the cross-sections
of photon scattering (elastic and inelastic), emission, absorption, inverse
bremsstrahlung, etc. and is a rather complicated task.
The energy transport
is assumed to be by radiation in the inner part of the Sun, and by convection
in the outer region. The border region between the radiative and convective zones
is located at $r \sim 0.7R_{\odot}$, where $r$ is the distance 
from the solar center. 

A solar model
should reproduce the observed physical characteristics of the Sun:
the mass \cite{AstrAlmanac} $M_{\odot} = (1.98892 \pm 0.00025)\times 10^{33}~{\rm g}$, 
the present radius $R_{\odot} = (6.9596 \pm 0.0007)\times 10^{5}~{\rm km}$
and luminosity \cite{BP95} (see also \cite{PD}) 
$L_{\odot} = 3.844(1 \pm 0.004)\times 10^{33}~{\rm erg~s^{-1}}$,
as well as the measured relative photospheric mass abundances of the elements
heavier than $^{4}$He, Z, and of the hydrogen, X: $(Z/X)_{photo} =
0.0245~(1 \pm 0.061)$; actually, these quantities are used as input in 
the relevant computer calculations.
An important constraint is the age of the Sun:
$\tau_{\odot} = (4.57 \pm 0.01)\times 10^{9}~{\rm yr}$. 
In order to develop a solar model one typically studies the evolution
of an initially homogeneous Sun, having a mass $M_{\odot}$ during a period of 
time  $\tau_{\odot}$. To reproduce the values of 
$R_{\odot}$, $L_{\odot}$ and $(Z/X)_{photo}$ at time $t = \tau_{\odot}$ 
three parameters in the calculations are used: 
the initial helium and heavy element abundances Y and Z, 
and a parameter characterizing the convection efficiency
in the outer region of the Sun (the mixing length parameter).
The latter is constrained by the value of $R_{\odot}$. 
It is usually assumed that the Sun is spherically symmetric.

It should be clear from the above brief discussion that the solar modeling
requires a rather good knowledge of several branches of physics:
astrophysics, atomic, nuclear 
(elementary particle) and plasma physics. The most recent and sophisticated
standard solar models 
\cite{Proff}--\cite{DS96}, \cite{HELIOSEIS1} 
include the effects of the slow
diffusion (relative to hydrogen) of helium and the heavier elements 
from the surface towards the center of the Sun (caused by the stronger 
gravitational pull of these elements relative to hydrogen).

\subsection{Helioseismological Constraints on Solar Models} 

At present one of the most stringent constraints on 
the solar models are obtained from the helioseismological data. 
It was discovered experimentally as early as in 1960
\cite{HELIOSEIS0} that the surface of the Sun is oscillating
with periods which vary in the interval between about 15 and 3 minutes
(these oscillations are usually referred  to as the 
``5 minute oscillations''). In later studies about
$10^{6}$ individual oscillation modes have been
identified experimentally and their frequencies  
were measured with an accuracy
of 1 part in $10^{4}$ or better. 

The Sun's surface oscillations reflect
the existence of standing pressure waves (p-waves) in the interior of the Sun
(see, e.g., \cite{HELIOSEIS0}).
Some of these waves penetrate deep into the region of neutrino production.
The p-mode frequencies depend on the physical 
conditions in the interior of the Sun.
Using a specially developed inversion technique and
the more precise helioseismological data on the low-frequency oscillations 
which became available recently, it was possible to reconstruct
with a remarkable accuracy
the sound speed distribution, $c(r)$, 
in a large region of the Sun
\cite{HELIOSEIS0,HELIOSEIS1,HELIOSEIS2}
extending from $r \cong 0.05~R_{\odot}$ to 
$r \cong 0.95~R_{\odot}$. Using the same data 
permitted to determine
the location of the bottom of the convective zone, $r_{b}$, and the matter density
at the bottom of the convective zone, $\rho_{b}$, as well
\cite{HELIOSEIS3}:
$$r_{b} = (0.708 - 0.714)~R_{\odot}~,~~~\eqno(7)$$  
$$\rho_{b} = (0.185 - 0.199)~{\rm g/cm^3}~.~~~\eqno(8)$$   

The implications of the helioseismological data for the solar modeling 
are illustrated in Fig. 4 (taken from \cite{HELIOSEIS2}), where the ratio
$(c_{SM}(r) - c_{HS}(r))/c_{HS}(r)$, $c_{HS}(r)$ and  $c_{SM}(r)$ being the 
sound speed distributions extracted from the helioseismological data
and predicted by a given solar model, is plotted for two solar models -
without and including heavy element diffusion \cite{BP95}.
Only the statistical errors in the determination 
of $c_{HS}(r)$ are shown (they are so small that they are barely seen), but the
general conclusions which can be inferred from such a comparison
remain valid after the inclusion
of conservatively estimated systematical errors \cite{HELIOSEIS3}.
As Fig. 4 illustrates, the difference between $c_{HS}(r)$ and $c_{SM}(r)$
for the model without heavy element diffusion is so large
that this model is practically ruled out by the helioseismological data.
Actually, the same conclusion is valid for all existing solar models
without heavy element diffusion (e.g., the models \cite{BP92,TCL,Cast1}). 

Further studies \cite{HELIOSEIS3}
have shown, in particular, that models which have been especially designed
to explain the observed deficiency of $^{8}$B neutrinos by lowering the
temperature in the central region of the Sun (see further), i.e., the so-called
models with ``mixed solar core'', also do not pass 
the helioseismological data test.
Thus, of the large number of solar models
proposed so far only the models which include diffusion of the heavy elements
are compatible with the helioseismological data.

\begin{figure}[ht]
\begin{center}
\epsfig{file=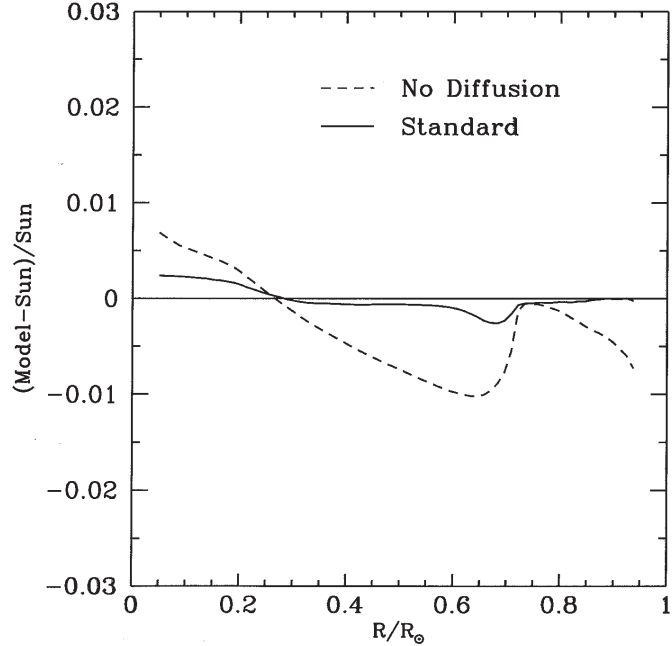,width=9cm}
\end{center}
\caption{The ratio
$(c_{SM} - c_{HS})/c_{HS}$, $c_{HS}$ and  $c_{SM}$ being the 
sound speed distributions extracted from the helioseismological data
and predicted by a given solar model, 
as a function of the distance $R$ from the center of the Sun  (expressed in
units of the solar radius ${\rm R_{\odot}}$) (from [36]
). The ratio
is plotted for two solar models --
without (dashed line) and including (solid line) 
heavy element diffusion [29]
. Only the statistical errors in the determination 
of $c_{HS}(r)$ are included in the analysis (they are too small to be seen).}
\end{figure}

The agreement of the predictions of the models with heavy element diffusion
for $c(r)$ with the sound speed distribution deduced from the data is quite
impressive. As Fig. 4 indicates, the root mean square (r.m.s.) deviation
from $c_{HS}(r)$ for the model of \cite{BP95} in the entire range of
$c_{HS}(r)$ determination is $\sim 0.1\%$. For the model with heavy element
diffusion of \cite{HELIOSEIS1} the r.m.s. discrepancy is $\sim 0.2\%$.
Adding the estimated systematic uncertainties in the determination of 
$c_{HS}(r)$ allows the difference
$|c_{SM}(r) - c_{HS}(r)|$ to be somewhat larger \cite{HELIOSEIS3}~:
the ratio $|c_{SM}(r) - c_{HS}(r)|/c_{HS}(r)$ for the model \cite{BP95}, 
however, does not exceed approximately 0.4\% in the region
$0.2R_{\odot} \leq r \leq 0.65R_{\odot}$, and about  
1\% in the neutrino production region  
$0.05 R_{\odot} \leq r \leq 0.20R_{\odot}$, for which the 
helioseismological data is less accurate.

As can be shown \cite{HELIOSEIS2}, in the interior of the Sun one has:
$c^{2}(r) \sim T(r)/\mu(r)$, where $T$ is the temperature and $\mu$ is the mean
molecular weight. Consequently, even small 
deviations of the solar model predictions
for $T$ and $\mu$ from their actual 
values in the Sun, $\delta T$ and $\delta \mu$, 
would lead to a relatively large discrepancy
between the predicted an measured values of 
$c(r)$: $\delta c /c \cong 0.5(\delta T /T - \delta \mu /\mu)$.
Since, according by the standard solar models, 
$T$ and $\mu$ vary by very different factors in the energy (neutrino) production
region,\footnote{They vary respectively by the factor of 53 and by 43\%
in the entire region of the $c(r)$ helioseismological determination.} 
namely, by a factor of 1.9 and by 39\%, it is quite unlikely that the
discrepancies between the predicted and actual values of     
$T$ and $\mu$ mutually cancel to produce the remarkable agreement
for the models with heavy element diffusion, illustrated in Fig. 4.
Barring such a cancellation, the comparison between   
$c_{HS}(r)$ and $c_{SM}(r)$ suggests that for the indicated models
$|\delta T|/T$, $|\delta \mu|/\mu \la 2\%$
in the interior of the Sun, and is considerably smaller in most part of the 
Sun. Since the solar neutrino fluxes scale approximately as
$T_{c}^{n}$ with $n = - 1.1;~8;~24$ respectively for 
the pp, $^{7}$Be and $^{8}$B
neutrinos \cite{BUlmer96}, $T_c$ being the 
central temperature predicted by the model, 
the above results implies, in particular, that it is 
impossible to reduce considerably
even the $^{8}$B neutrino flux by changing the central temperature
within the limits following from the helioseismological data.

\subsection{Predictions for the Neutrino Fluxes and the Signals
in the Solar Neutrino Detectors}

We shall present here the results obtained in four models 
\cite{BP95}--\cite{DS96}  
with heavy element diffusion, 
which can be characterized by their predictions for the total flux of $^{8}$B
neutrinos as relatively ``high flux'' \cite{BP95,RVCD}, 
``intermediate flux'' \cite{Cast2} 
and ``low flux'' \cite{DS96} models. 
The predictions for the $^{8}$B neutrino flux and for 
the signals in the solar neutrino detectors in these models determine the 
corresponding intervals in which the results of practically all 
contemporary solar models compatible with the helioseismological
data 
\cite{HELIOSEIS01}--\cite{HELIOSEIS3} 
lie (see, e.g.,  \cite{Proff,HELIOSEIS1}). 
Thus, they give an idea about the dispersion and the possible
uncertainties in the predictions. 

In Table \ref{tab:effluents1} we have collected the results of the models of 
Bahcall--Pinsonneault (BP'95) from 1995 \cite{BP95},  
Richard et al. (RVCD) \cite{RVCD}, 
Castellani et al. (CDFLR) \cite{Cast2},
and of Dar and Shaviv from 1996 (DS'96) \cite{DS96} 
for the values of the fluxes of the 
pp, $^{7}$Be, $^{8}$B, pep and CNO neutrinos at the Earth surface. We have
included also the estimated 1 s.d. uncertainties in the predictions for the
fluxes made by Bahcall and Pinsonneault for 
their model. In Tables \ref{tab:effluents2} and \ref{tab:effluents3} we give the
predictions for the contributions of each of the indicated six fluxes to the
signals in the Cl--Ar \cite{Davis} and the Ga--Ge \cite{SAGE,GALLEX} 
experiments, respectively, and
quote the predictions for the total signals in these experiments (including the
estimated 1 s.d. uncertainty in the predictions whenever it is given by the
authors).

A comparison between the experimental results (1) -- (6) and the
corresponding predictions given in Tables \ref{tab:effluents2} and \ref{tab:effluents3} leads to the conclusion
that none of the solar models proposed so far provides a satisfactory 
description of the solar neutrino data: the predictions 
typically exceed the observations. This is
one of the current aspects of the solar neutrino problem.

\begin{table*}[hbt]
\setlength{\tabcolsep}{1.5pc}
\newlength{\digitwidth}
\settowidth{\digitwidth}{\rm 0}
\catcode`?=\active \def?{\kern\digitwidth}
\caption{Solar neutrino fluxes at the Earth surface 
(in units of cm$^{-2}$sec$^{-1}$) predicted by the solar models [29]--[32].}
\label{tab:effluents1}
\begin{tabular*}{\textwidth}{@{}l@{\extracolsep{\fill}}rrrr}
\hline
Flux\phantom{ of neutrinos}&{BP'95}
                 &{RVCD}
                 &{CDFLR}
                 &{DS'96}         \\
\hline\vspace{-3mm}\\
$\Phi_{\rm pp}\times 10^{-10}$ & $ 5.91(1~^{+0.01}_{-0.02})$ & $5.94$ & $ 5.99$ & 
$6.10$\smallskip\\
$\Phi_{\rm pep}\times 10^{-8}$ & $1.40(1~^{+0.01}_{-0.02})$ & $1.38$ & $1.40$ & 
$1.43$\smallskip\\
$\Phi_{\rm Be}\times 10^{-9}$ & $5.15(1~^{+0.06}_{-0.07}) $ & $4.80$ & $ 4.49$ &
$3.71$\smallskip\\
$\Phi_{\rm B}\times 10^{-6}$ & $6.62(1~^{+0.14}_{-0.17}) $ & $6.33$ & $5.16$ &
$2.49$\smallskip\\
$\Phi_{\rm N}\times 10^{-8}$ & $6.18(1 ~^{+0.17}_{-0.20}) $ & $5.59$ & $5.30$ & 
$3.82$\smallskip\\
$\Phi_{\rm O}\times 10^{-8}$ & $5.45(1~^{+0.19}_{-0.22}) $ & $4.81$ & $4.50$ & 
$3.74$\smallskip\\
\hline
\multicolumn{5}{@{}p{120mm}}{}
\end{tabular*}
\end{table*}

\begin{table*}[hbt]
\setlength{\tabcolsep}{1.5pc}
\settowidth{\digitwidth}{\rm 0}
\catcode`?=\active \def?{\kern\digitwidth}
\caption{Signals (in SNU) in Cl--Ar detectors due to the solar neutrinos,
predicted by the solar models [29]--[32].}
\label{tab:effluents2}
\begin{tabular*}{\textwidth}{@{}l@{\extracolsep{\fill}}rrrr}
\hline
Type of neutrinos             &{BP'95}
                              &{RVCD}
                              &{CDFLR}
                              &{DS'96}         \\
\hline
pp         & $ 0.0$ & $0.0$ & $ 0.0$ & $0.0$\\
pep & $ 0.22$ & $0.22$ & $0.22$ & $0.24$\\
$^{7}$Be & $1.24$ & $1.15$ & $ 1.08$ & $0.89$\\
$^{8}$B & $7.36$ & $6.71$ & $5.74$ & $2.64$ \\
$^{13}$N & $0.11$ & $0.09$ & $0.09$ & $0.06$\\
$^{15}$O & $0.37$ & $0.32$ & $0.31$ & $0.25$\\
\hline\vspace{-3mm}\\
Total:    & $9.3~^{+1.2}_{-1.4}$ & $8.5$ & $7.4$ & $4.1~\pm~1.2$\smallskip\\
\hline
\multicolumn{5}{@{}p{120mm}}{}
\end{tabular*}
\end{table*}

Taking into account the estimated uncertainties in the theoretical 
predictions and the experimental errors in  (1) -- (6) one finds that 
the differences between the predictions and the observations are
largest for the ``high flux'' models: the measured values of 
$\bar{\rm R}$(Ar), $\bar\Phi$(B) and $\bar {\rm R}({\rm Ge})$ in the Cl--Ar, 
Kamiokande and Ga--Ge experiments are at least by (3.5--4.0) s.d. smaller than
the predicted ones in \cite{BP95,RVCD}. The ``low flux'' model of Dar and Shaviv 
reproduces the result of the Kamiokande and Super-Kamiokande 
experiments for $\bar\Phi$(B). 
However, the prediction of the model for $\bar {\rm R}({\rm Ge})$ is 
respectively by at least 3 s.d. 
higher than the GALLEX and SAGE results (3) and (4). The discrepancy between
the solar model predictions and the Ga--Ge data is somewhat larger if one
compares the predictions with the combined result (5) of the GALLEX and SAGE
experiments. 

\begin{table*}[hbt]
\setlength{\tabcolsep}{1.5pc}
\settowidth{\digitwidth}{\rm 0}
\catcode`?=\active \def?{\kern\digitwidth}
\caption{Signals (in SNU) in Ga--Ge detectors due to the solar neutrinos,
predicted by the solar models [29]--[32].}
\label{tab:effluents3}
\begin{tabular*}{\textwidth}{@{}l@{\extracolsep{\fill}}rrrr}
\hline

Type of neutrinos             & \multicolumn{1}{r}{BP'95}
                              & \multicolumn{1}{r}{RVCD}
                              & \multicolumn{1}{r}{CDFLR}
                              & \multicolumn{1}{r}{DS'96}         \\
\hline
pp         & $ 69.7$ & $70.1$ & $ 70.7$ & $72.0$ \\
pep & $3.0$ & $3.0$ & $3.0$ & $3.0$ \\
$^{7}$Be & $37.7$ & $35.3$ & $32.9$ & $27.1$ \\
$^{8}$B & $16.1$ & $15.4$ & $12.6$ & $6.1$ \\
$^{13}$N & $3.8$ & $3.5$ & $3.3$ & $2.4$ \\
$^{15}$O & $6.3$ & $5.6$ & $5.2$ & $4.6$ \\
\hline\vspace{-3mm}\\
Total:    & $137~^{+8}_{-7}$ & $133$ & $128$ & $115~\pm~6.0$\smallskip\\
\hline
\multicolumn{5}{@{}p{120mm}}{}
\end{tabular*}
\end{table*}

Let us discuss in somewhat greater detail the results of the four
representative models \cite{BP95}--\cite{DS96}  
for the fluxes 
of the pp, pep, $^{7}$Be, $^{8}$B
and CNO neutrinos shown in Table \ref{tab:effluents1}. The predictions for the values of the pp
and pep neutrino fluxes are remarkably coherent: they very from model to model
at most by 3\% and 3.5\%, respectively. Actually, the two fluxes are related
\cite{JNB}: $\Phi_{\rm pep}$ is proportional 
to $\Phi_{\rm pp}$ and the coefficient
of proportionality is practically solar model independent, being determined by
the ratio of the cross--sections of the reactions ${\rm p + e^{-} + 
p \rightarrow D + \nu_e}$ and ${\rm p + p \rightarrow D + e^{+} + \nu_e}$ in 
which the pep and pp neutrinos are produced in the Sun. One has 
\cite{JNB}--\cite{DS96}  
$$ \Phi_{\rm pep} = (2.3 - 2.4)\times 10^{-3} \Phi_{\rm pp}.~~~\eqno(9) $$

The value of the pp flux is constrained by the existing
rather precise data on the solar luminosity, L$_{\odot}$. Indeed \cite{Lumin}, the solar
luminosity is determined by the thermal energy released in the Sun in the two
well known cycles of nuclear reactions, the pp (Fig. 1) and the CNO cycles 
(see, e.g., \cite{JNB}), in which
4 protons are converted into $^{4}$He with emission of 2 neutrinos. From the
point of view of the energy effectively generated, the indicated hydrogen
burning reactions can be written as
$$ 4{\rm p} + 2{\rm e}^{-} \rightarrow~^{4}{\rm He} + 2\nu_e.~~\eqno (10)$$
Depending on the cycle, the two emitted neutrinos can be both of the
pp or pep type, or a pp (pep) and a $^{7}$Be, a pp (pep) and a $^{8}$B (pp
cycle), and a $^{15}$O and/or $^{13}$N (CNO cycle) neutrinos \cite{JNB,Cast2}. 
The thermal energy released per one produced pp, pep, $^{7}$Be and $^{8}$B neutrino is
equal to $({\rm Q}/2 - \bar{\rm E}_{{\rm j}})$, where Q $=$ 26.732 MeV is the
Q--value of the reaction (7), and $\bar{\rm E}_{{\rm j}}$ is the average energy
of the neutrino of the type j (j = pp, ...). The energy released per one 
$^{15}$O and/or one $^{13}$N neutrino, as can be shown (taking into 
account, in particular, the discussion of the 
rates of the different reactions of the CNO cycle given in \cite{JNB})
is equal with a high precision to the difference 
$({\rm Q}/2 - (\bar{\rm E}_{{\rm N}} + \bar{\rm E}_{{\rm O}})/2)$. Obviously, 
the values of
Q and $\bar{\rm E}_{{\rm j}}$ are solar physics (and therefore solar model)
independent. 

Given the average energies $\bar{\rm E}_{{\rm j}}$ carried away 
by the pp, pep, $^{7}$Be, $^{8}$B and CNO neutrinos (they are listed at the 
beginning of Sect. 2, $\bar{\rm E}_{{\rm Be}} = 0.813~$MeV), and knowing 
that the energy emission by the Sun is quasi-stationary (steady state), it is 
possible to
relate L$_{\odot}$ with the pp, pep, $^{7}$Be, $^{8}$B and CNO neutrino
luminosities of the Sun. One finds in this way the following constraint on the
solar neutrino fluxes:
$$
\Phi_{\rm pp} + 0.958\,\Phi_{\rm Be} + 0.955 \,\Phi_{\rm CNO} 
+ 0.910\,\Phi_{\rm pep}= (6.514~\pm~0.031) 
\times 10^{10}~{\rm cm}^{-2}{\rm sec}^{-1}~,~~\eqno(11)
$$
where 
$$
\Phi_{\rm CNO} = \Phi_{\rm N} + \Phi_{\rm O}\, ,$$ 
we have used 
\cite{BP95} (see also \cite{AstrAlmanac,PD}) 
L$_{\odot} = (3.844~\pm~0.015) \times 10^{33}~{\rm erg}~{\rm sec}^{-1}$  
and have neglected the terms proportional to $\Phi_{\rm B}$ and to  
$(\Phi_{\rm N} - \Phi_{\rm O})$ in the left hand side of the 
equation, which are predicted to be considerably 
smaller than $3\times 10^{8}~{\rm cm}^{-2}{\rm sec}^{-1}$.
The fluxes $\Phi_{\rm B}$ and $(\Phi_{\rm N} - \Phi_{\rm O})$ have 
to be more than 46 and (3 - 4) times bigger than the largest (``high
flux'') model predictions \cite{BP95} and \cite{RVCD}, respectively, 
in order for these terms 
to exceed the indicated value. The coefficient multiplying the 
$\Phi_{\rm Be}$ term in  (11) is just the ratio of the thermal energies 
produced per one $^{7}$Be and one pp neutrino,
$({\rm Q}/2 - \bar{\rm E}_{{\rm Be}})$/$({\rm Q}/2 - \bar{\rm E}_{{\rm pp}})$,
etc. Since, as Table \ref{tab:effluents1}
shows, $\Phi_{\rm Be}$ and  $\Phi_{\rm CNO}$ are smaller than $\Phi_{\rm pp}$
at least by the factors 0.09 and 0.02, respectively, and $\Phi_{\rm pep}$ is
even smaller (see  (9)),  (11) limits primarily the pp neutrino flux.

Comparing the experimental results (3) -- (6) with the solar model
predictions given in Table \ref{tab:effluents3} one notices, in particular, that the rate of 
$^{71}$Ge production due only to the pp neutrinos, R$^{{\rm pp}}({\rm Ge}) = 
(70-72)~$SNU, is very close to the rates observed in the GALLEX and SAGE
experiments. This suggests that a large fraction, i.e., roughly at 
least half, of the pp (electron) neutrinos emitted by the Sun reach the 
Earth intact and are detected. 

The relative spread in the predictions for the $^{7}$Be neutrino flux 
$\Phi_{\rm Be}$ and for the signals due to the $^{7}$Be 
neutrinos in the Cl--Ar
and Ga--Ge experiments, as it follows from Tables \ref{tab:effluents1} -- 
\ref{tab:effluents3}, does not exceed 
approximately 30\%. 

\section{The $^{8}$B Neutrino Problem}     

A further inspection of the results collected in Table \ref{tab:effluents1} reveals that the
differences (and the estimated uncertainties) in the predictions of the solar
models for the total flux of $^{8}$B neutrinos are the largest: the value of
$\Phi_{\rm B}$ in the models \cite{BP95}--\cite{Cast2}  
is by more than a factor of 2
larger than the value obtained in the ``low flux'' model \cite{DS96}. The $^{8}$B 
neutrinos are born in the Sun in the $\beta^{+}$-decay, 
$^{8}{\rm B} \rightarrow~^{8}{\rm Be}^{*} + {\rm e}^{+} + \nu_e$, of the
$^{8}$B nucleus which is produced in the reaction 
$$ {\rm p} +~^{7}{\rm Be} \rightarrow~^{8}{\rm B} + \gamma ~~~\eqno(12) $$
initiated by $\sim$20 keV protons. Obviously, $\Phi_{\rm B}$ is proportional 
to the rate of the process (12) taking place in the solar plasma environment, 
which in turn is to large extent determined by the cross--section of (12), 
$\sigma_{{\rm 17}}({\rm E_p})$. The latter is usually represented in the form
\cite{JNB}
$$
\sigma_{{\rm 17}}({\rm E_p}) = {{\rm S_{17}(E_p)}\over {{\rm E_p}}}~
{\rm exp (-8\pi~e^2/v)},~~~\eqno(13)
$$
where ${\rm exp (-8\pi~e^2/v)}$ is the Gamow penetration factor, and 
E$_{\rm p}$ and v are the p -- $^{7}$Be c.m. kinetic energy and relative 
velocity. The largely
different values of the astrophysical factor ${\rm S_{17}}$, ${\rm S_{17}}
\equiv {{\rm S_{17}({\rm E_p \sim 20~ keV})}}$, adopted by the authors of  
\cite{BP95,RVCD,Cast2} and of \cite{DS96} is one of the major 
sources of the large spread in the 
predictions for $\Phi_{\rm B}$. 
 
Because of background problems it is impossible to measure the
cross--section $\sigma_{{\rm 17}}({\rm E_p})$ directly at the low energies of
the incident protons, which are of astrophysical interest. The experimental 
studies of the process (12) were performed at energies 
$110~{\rm keV} \leq {\rm E_p} \la 2000~$keV. They are technically rather
difficult because of the instability of the $^{7}$Be serving as a target. The
results obtained in the indicated higher energy domain are extrapolated to 
${\rm E_p} \sim 20~$keV using a theoretical model describing the data (and the 
process (12) in the entire energy range 
$20~{\rm keV} \leq {\rm E_p} \la 2000~$keV) and taking into account the
possible solar plasma screening effects. Obviously, there are at least two
major sources of uncertainties in the determination of ${\rm S_{17}}$ inherent
to the indicated approach: the uncertainties associated with the data at 
${\rm E_p} \geq 110~$keV, and those associated with the extrapolation
procedure exploited.

Altogether six experiments have provided data on the
the p -- $^{7}$Be cross--section $\sigma_{{\rm 17}}({\rm E_p})~$ so far. 
The results of the four most accurate of them \cite{Kavanagh,Filip} can be
grouped in two distinct pairs, \cite{Kavanagh} and \cite{Filip}, 
which agree on the energy 
dependence of ${\rm S_{17}(E_p)}$, but disagree systematically by 
$~\sim$ (20--25)\% ($~\sim$(2--3) s.d.) on the absolute values of 
${\rm S_{17}(E_p)}$. The authors of \cite{BP95,Cast2} (and, we suppose, of 
\cite{RVCD}) used in their
calculations the value 
${\rm S_{17}} = (22.4~\pm~2.1)~$eV-b derived by extrapolation in  
\cite{John} on the basis of the data from all six experiments. 

A new method of experimental determination of ${\rm S_{17}}$ was proposed
relatively recently in \cite{Moto}. It is based on 
the idea of measuring the cross-section
of the inverse reaction,  $\gamma +~^{8}$B~$\rightarrow~^{7}$Be$~+~$p, by 
studying the dissociation of $^{8}$B into p + $^{7}$Be in the
Coulomb field of a heavy nucleus, chosen to be $^{208}$Pb. The 
time--reversal symmetry guarantees that the 
cross--sections of (12) and of the inverse reaction should be equal. 
The extraction of the values of 
$\sigma_{{\rm 17}}({\rm E_p})$ (and of ${\rm S_{17}(E_p)}$) from the data on 
the
process $^{8}{\rm B +~^{208}Pb} \rightarrow {\rm p} +~^{7}{\rm Be +~^{208}Pb}$ 
is not straightforward and is associated with certain subtleties 
(see, e.g., \cite{Lang}). 

Using the results of the experiment of Motobayashi et al.
\cite{Moto} on the reaction $^{8}{\rm B + ~^{208}Pb} 
\rightarrow {\rm p}~ + ~^{7}{\rm Be + ~^{208}Pb}$
to determine the cross-section $\sigma_{{\rm 17}}({\rm E_p})$ in the energy 
interval $500~{\rm keV} \la {\rm E_p} \la 2000~$keV, the results of the
most recent of the experiments on (12) of Filippone et al. \cite{Filip}
in the interval (110 -- 500) keV, and a new extrapolation model developed by
them, the authors of \cite{DS96} obtain ${\rm S_{17}} = 17 \pm 2~$eV-b.~%
\footnote{Let us note that
in  \cite{TCL} the value of ${\rm S_{17}}$ derived in 
\cite{John} and adopted in  \cite{BP95,Cast2} 
was also used, but with a larger 
systematical error, ${\rm S_{17}} = (22.4~\pm~1.3~ 
\pm~3.0)~$eV-b, introduced to account for  the (systematic) difference between 
the data on $\sigma_{{\rm 17}}({\rm E_p})$ from the experiments 
\cite{Kavanagh} and \cite{Filip}.}  
   
The additional difference between the values of $\Phi_{\rm B}$ predicted 
in \cite{BP95} and in \cite{DS96} is due to
\begin{enumerate}
\item[i)] the use of different 
(but still compatible within the errors with the measured
or deduced from the data)
values of other relevant
nuclear reaction cross--sections, as those of the 
$^{3}{\rm He +~^{3}He} \rightarrow~^{4}{\rm He + 2p}$ and 
$^{3}{\rm He +~^{4}He} \rightarrow~^{7}{\rm Be} + \gamma~$ 
reactions (a factor $\sim 1.2$), and 
\item[ii)] the use in  \cite{BP95} and in  \cite{DS96} of different methods to
account for the diffusion of the heavy elements in the Sun. 
\end{enumerate}
The latter leads, in particular, to a difference in the values
of the central temperature in the Sun in the models
\cite{BP95} and \cite{DS96}: 
${\rm T_{c}(BP'95) = 1.584\times 10^{7}~K}$ and 
${\rm T_{c}(DS'96) = 1.561\times 10^{7}~K}$. As the $^{8}$B neutrino flux 
$\Phi_{\rm B}$ is very
sensitive to the value of T$_{\rm c}$, scaling as \cite{BUlmer96}
$\Phi_{\rm B} \sim {\rm T}^{24}_{\rm c}~$, 
the indicated difference in T$_{\rm c}$ implies an additional difference in the
values of $\Phi_{\rm B}$ predicted in \cite{BP95} and in \cite{DS96} 
(a factor of $\sim 1.4$).

\section{The Missing $^{7}$Be Neutrinos}

Even if one accepts  
that there are large uncertainties in the predictions for
the flux of $^{8}$B neutrinos and in all analyses one should 
rather use for $\Phi_{\rm B}$ the value 
implied by the (Super{-}) Kamiokande data, 
(2) and (6), another problem arises: 
the predictions of all contemporary solar models 
for the flux of $^{7}$Be neutrinos, $\Phi_{\rm Be}$, are considerably larger 
than the value suggested by the existing solar neutrino data.
This was first noticed in  \cite{BachBe} and confirmed in several subsequent more 
detailed studies \cite{Cast3}--\cite{Bach2}  
utilizing a variety of 
different methods. We shall illustrate here the above result using rather simple 
arguments \cite{BachBe,Bach2}.
  
Let us assume that the spectrum of the $^{8}$B neutrino flux coincides with
that predicted by the solar models, i.e., with the spectrum of 
the $\nu_e$ emitted in the decay 
$^{8}{\rm B} \rightarrow~^{8}{\rm Be}^{*} + {\rm e}^{+} + \nu_e$ (see Fig. 2).
This would be the case if the solar $^{8}$B $\nu_e$ behave conventionally during their 
journey to the Earth. For the value of the total  $^{8}$B neutrino flux one can use the 
Super-Kamiokande result (6):
$\bar\Phi^{SK}_{{\rm B}} = (2.44~^{+0.26}_{-0.11}) 
\times 10^{6}~{\rm cm}^{-2}{\rm sec}^{-1}$, where the statistical and the systematic
errors were added in quadratures. Knowing ${\rm \Phi_{B}}$ and the  
the cross-section \cite{JNB} of the Pontecorvo--Davis reaction 
$\nu_e + ^{37}{\rm Cl} \rightarrow e^{-} + ^{37}{\rm Ar}$, 
one can calculate the contribution of
the $^{8}$B neutrinos to the signal in the Davis et al. experiment, 
${\rm R^{B}(Ar)}$. One finds:
$${\rm R^{B}(Ar)} = (2.71~^{+0.30}_{-0.14})~{\rm SNU}.~~\eqno(14)$$
By subtracting this value from the rate of Ar production
measured in the Davis et al. experiment we obtain 
the sum of the contributions of the $^{7}$Be, pep and CNO neutrinos
to the signal in this experiment:
$${\rm R^{Be+pep+CNO}(Ar)} = ~(-0.15~^{+0.37}_{-0.25})~{\rm SNU}.~~~\eqno(15)$$
Given the solar model independent relation (9) between 
${\rm \Phi_{pep}}$ and ${\rm \Phi_{pp}}$, and that ${\rm \Phi_{pp}}$ is rather
tightly constrained by the data on the solar luminosity, we can consider as
rather reliable (and weakly model dependent) the solar model predictions for 
${\rm R^{pep}_{SM}(Ar)} = (0.22 - 0.24)~$SNU. Taking 
${\rm R^{pep}(Ar)} = 0.22~$SNU one obtains from (15)
$$ {\rm R^{Be}(Ar)} \leq ~(-0.37~^{+0.37}_{-0.25})~{\rm SNU},~~~\eqno(16)$$
$ {\rm R^{Be}(Ar)}$ being the $^{7}$Be neutrino contribution 
to the signal in the Davis et al. experiment.
At 99.73\% C.L. (3 s.d.) this implies 
${\rm R^{Be}(Ar)} \leq 0.74~{\rm SNU}$, which is smaller than 
the predictions of the solar models with heavy element 
diffusion,\footnote{Actually, the 3 s.d. upper limit on ${\rm R^{Be}(Ar)}$ is smaller
than the predictions of all known to the author 
solar models proposed in the last 10 years (see also the second article quoted
in  \cite{HELIOSEIS2}).}
${\rm R^{Be}_{SM}(Ar)} = (0.89 - 1.24)~{\rm SNU}~$
\cite{Proff,BP95,RVCD,Cast2,DS96,HELIOSEIS1}. 
As $ {\rm R^{Be}(Ar)} \sim \Phi_{\rm Be}$, the result obtained suggests
that, if the solar neutrinos are assumed to behave conventionally 
on the way to the Earth (i.e., do not 
undergo oscillations, transitions, decays, etc.),
the $^{7}$Be $\nu_e$ flux inferred from the solar neutrino data 
is substantially smaller than the flux 
predicted by the contemporary solar models.

Similar (though statistically somewhat weaker) conclusions can be reached
for the contribution of the $^{7}$Be neutrinos to the signal 
in the Ga-Ge detectors, ${\rm R^{Be}(Ge)}$, and correspondingly for 
$\Phi_{\rm Be}$, by taking into account the fact
that the solar model predictions for the contributions 
of the pp and pep neutrinos
to the indicated signal, ${\rm R^{pp+pep}(Ge)}$, are tightly 
constrained by the data on the solar luminosity
and vary by no more than 3\%: ${\rm R^{pp+pep}_{SM}(Ge)} = (72.7 - 75.0)~{\rm SNU}$.   
Subtracting ${\rm R^{pp+pep}_{SM}(Ge)} = 72~{\rm SNU}$ from the 
rate of Ge production observed in 
the SAGE and GALLEX experiments, Eq. (5), one obtains
for the contribution of $^{8}$B, $^{7}$Be and CNO neutrinos:
${\rm R^{B+Be+CNO}(Ge)} = (3.4~ \pm~ 7)~{\rm SNU}$.   

Utilizing the  value of
$\bar{\Phi}^{SK}_{{\rm B}}$ measured by the Super-Kamiokande experiment 
and the Ga--Ge reaction 
cross--section \cite{JNB,Bach2} permits to calculate the contribution of the
$^{8}$B neutrinos, ${\rm R^{B}(Ge)}$, to ${\rm \bar{R}_{exp}(Ge)}$:
${\rm R^{B}(Ge)} = (5.9~^{+5.9}_{-2.9})~{\rm SNU}$, where the error is dominated
by the estimated uncertainty in the value of the Ga--Ge reaction 
cross--section \cite{Bach2}. Subtracting the so derived 
value of ${\rm R^{B}(Ge)}$ 
from the value of ${\rm R^{B+Be+CNO}(Ge)}$ we get:
$$ {\rm R^{Be}(Ge)} \leq ~(-2.5~^{+9.2}_{-7.6})~{\rm SNU}.~~~\eqno(17)$$
Consequently, at 99.73\% C.L. (3 s.d.) 
the contribution of $^{7}$Be neutrinos
to the signals in the SAGE and GALLEX experiments does not exceed
25.1 SNU, while the solar models 
\cite{BP92}--\cite{DS96}, \cite{HELIOSEIS2} 
predict ${\rm R^{Be}(Ge)} \geq 27~{\rm SNU}$.

Analogous results have been obtained in  \cite{Cast3,KwRos} using different
methods. The same conclusion has been reached 
in \cite{HatLang1} as well on the basis of 
a $\chi^2-~$analysis of the solar model 
description of the data, in which the total pp, 
pep, $^{7}$Be, $^{8}$B and CNO
neutrino fluxes were treated as free parameters subject only to the luminosity
constraint (11), while the spectra of solar neutrinos were assumed to coincide
with the predicted ones in the absence of unconventional neutrino behaviour
(as oscillations in vacuum, etc.). 
    
Thus, there are strong indications from the existing solar
neutrino data that the flux of $^{7}$Be (electron) neutrinos is considerably
smaller than the flux predicted in all contemporary solar models. 
Given the results of the GALLEX and SAGE calibration experiments, 
we can conclude that both the Davis et
al. and the Super-Kamiokande (Kamiokande) data,  (2) and  (6) (Eq. (3)),  
have to be incorrect in order for the
above conclusion to be not valid. The discrepancy bet\-ween the value of 
$\Phi_{{\rm Be}}$ suggested by the analyses of the existing solar neutrino 
data and the solar model predictions for $\Phi_{{\rm Be}}$ represents the 
major new aspect of the solar neutrino problem. No plausible astrophysical 
and/or nuclear physics explanation of this discrepancy has been proposed so far.

\section{Neutrino Physics Solutions of the Solar Neutrino Problem}

We have seen that none of the solar models developed during the last ten years
provides a satisfactory description of the existing solar neutrino data.
The discrepancy between the data and the solar model predictions is
especially large for the majority of models with heavy element diffusion, which are
compatible with the helioseismological data. 
The solar model predictions for the signals caused by the solar neutrinos 
in the solar neutrino experiments are larger
than the measured signals. In particular, no solar, atomic 
or nuclear physics solution to the  
$^{7}$Be neutrino problem discussed above was found so far.
Since the solar neutrino detectors 
are sensitive either only,
or predominantly, to the solar $\nu_e$ flux, these results 
indicate that the solar $\nu_e$ flux
is depleted on the way to the Earth. 

Such a depletion can take place
naturally if the solar $\nu_e$  
undergo transitions into neutrinos of a different type, $\nu_{\mu}$ and/or
$\nu_{\tau}$, and/or into a sterile neutrino $\nu_{s}$, or are 
converted into antineutrinos $\bar{\nu}_{\mu}$ and/or $\bar{\nu}_{\tau}$,
while they travel to the Earth. The depletion of the solar $\nu_e$ flux might be 
caused also by instability of the solar neutrinos which can decay on their way
to the Earth. Thus, several physically rather different neutrino physics solutions
of the solar neutrino problem are, in principle, possible. They all require
the existence of ``unconventional''
intrinsic neutrino properties (mass, mixing, magnetic moment) and/or
couplings (e.g., flavour changing neutral current (FCNC)
interactions). More specifically, these solutions 
include:
\begin{enumerate}
\item[i)] oscillations in vacuum \cite{Pont57,Pont67,NMP2} of the
solar $\nu_e$ into different weak eigenstate neutrinos ($\nu_{\mu}$
and/or $\nu_{\tau}$, and/or sterile neutrinos, $\nu_s$) on the way
from the surface of the Sun to the Earth \cite{KP1}--\cite{BerRos}, 
\item[ii)] matter-enhanced transitions
\cite{MSW1,MSW2} $\nu_e \rightarrow \nu_{\mu(\tau)}$, and/or
$\nu_e \rightarrow \nu_s$, while the solar neutrinos propagate from the
central part to the surface of the Sun \cite{KP4}--\cite{KP5},  
\item[iii)] solar $\nu_e$ resonant spin or spin-flavour conversion (RSFC) in the
magnetic field of the Sun \cite{RSFP1}, and
\item[iv)] matter-enhanced transitions,
for instance $\nu_e \rightarrow \nu_{\tau}$, in the Sun, induced by 
flavour changing neutral current (FCNC)
interactions of the solar $\nu_e$ with the particles forming the solar
matter \cite{FCNC1}--\cite{FCNC3}  
(these transitions can take place even in the case of
absence of lepton mixing in vacuum and massless neutrinos \cite{FCNC1}). 
\end{enumerate}
All these possibilities have been and continue to be extensively studied 
(see the quoted articles).  
There have not been recent studies of the solar neutrino 
decay hypothesis \cite{DEC1} which, however, 
was disfavored \cite{DEC2} by the earlier solar neutrino data.

In what follows we shall discuss the vacuum oscillation and 
the matter-enhanced transition solutions of the solar neutrino problem. The status of 
these solutions has been reviewed recently, e.g., in \cite{SPnu96,SPHeid97}.

\subsection{Oscillations in Vacuum}

Neutrino oscillations in vacuum \cite{Pont57}, have
been discussed in connection with the solar neutrino experiments
\cite{Pont67} and as a possible solution of the solar neutrino 
problem \cite{NMP2,NMP1}, \cite{KP1}--\cite{BerRos}, \cite{SPHeid97}  
(and the literature quoted therein) 
for about 31 years. In the simplest version of this 
scenario it is assumed that the state
vector of the electron neutrino, $|\nu_e\rangle$, produced in vacuum with momentum
$\vectorp$ in some weak interaction process, 
is a coherent superposition of the state vectors $|\nu_i\rangle$ of two neutrinos 
$\nu_i$, i=1,2, having the same momentum $\vectorp$
and definite but different masses in vacuum, 
${\rm m_i}$, ${\rm m_1} \neq {\rm m_2}$, while the linear 
combination of $|\nu_1\rangle$ and $|\nu_2\rangle$, which is 
orthogonal to $|\nu_e\rangle$, represents the state vector $|\nu_x\rangle$ of another
weak-eigenstate neutrino, $|\nu_x\rangle~ = |\nu_{\mu (\tau)}\rangle$ or $|\nu_{s}\rangle$: 
$$|\nu_e\rangle~ = |\nu_1\rangle \cos\theta + |\nu_2\rangle \sin\theta~,~~~~\eqno(18a)$$
$$|\nu_x\rangle~ = - |\nu_1\rangle \sin\theta + |\nu_2\rangle \cos\theta~,~~~\eqno(18b)$$
where $\theta$ is the neutrino (lepton) mixing angle in vacuum. 
We shall assume for concreteness in what follows that 
$\nu_x$ is an active neutrino, say $\nu_{\mu}$,
$|\nu_x\rangle~ = |\nu_{\mu}\rangle$. 

Obviously, the states $|\nu_{1,2}\rangle$ are
eigenstates of the Hamiltonian of the neutrino system 
in vacuum, $H_{0}$:
$$ H_{0}~|\nu_i\rangle~ = E_{i}~|\nu_i\rangle,~~~ E_{i} = 
\sqrt{\vectorp^2 + m_{i}^2},~i=1,2.~~~\eqno(19)$$

If $\nu_e$ is produced
at time $t = 0$ in the Sun in the state given by (18a), 
after a time $t$ the latter will evolve into the state 
$$|\nu_e(t)\rangle~ = e^{-iE_{1}t}~|\nu_1\rangle \cos\theta +  e^{-iE_{2}t}~|\nu_2\rangle \sin\theta~,
\eqno(20)$$
where we have ignored the overall 
space coordinate dependent factor ${\rm exp}(i\,\vectorp\,\vectorr)$ in 
the right-hand side of (20) and have assumed that 
the solar matter does not affect the evolution of the neutrino system. 
(The possible effects of matter
on the evolution of the neutrino state will be 
considered in the next Section.)
Using (18a) and (18b) to express 
the vectors $|\nu_1\rangle$ and $|\nu_2\rangle$ in terms of the vectors
$|\nu_{e}\rangle$ and $|\nu_{\mu}\rangle$ we can rewrite (20) in the form:
$$
|\nu_e(t)\rangle~ = A_{ee}(t)~|\nu_e\rangle + 
A_{\mu e}(t)~|\nu_{\mu}\rangle~,~~~\eqno(21)
$$ 
where 
$$
A_{ee}(t) = e^{-iE_{1}t}~\cos^2\theta + 
e^{-iE_{2}t}~\sin^2\theta~~\eqno(22)
$$ 
and 
$$
A_{\mu e}(t) = \frac{1}{2}~\sin 2\theta~ 
(e^{-iE_{2}t} - e^{-iE_{1}t})~~\eqno(23)
$$ 
are the probability amplitudes to find respectively 
neutrino $\nu_e$ and neutrino $\nu_{\mu}$ at time t of
the evolution of the neutrino system 
if neutrino $\nu_e$ has been produced at time $t = 0$.
Thus, if neutrinos $\nu_1$ and $\nu_2$ are not mass-degenerate,
$m_1 \neq m_2$, and if nontrivial neutrino mixing exists in vacuum, $\theta \neq
n\pi /2$, $n=0,1,2,...$, we have 
$ |A_{\mu e}(t)|^2 \neq 0$ and 
transitions in flight between the
states $|\nu_e\rangle$ and $|\nu_{\mu}\rangle$  (i.e., between the neutrinos
$\nu_e$ and $\nu_{\mu}$) are possible.

Assuming that neutrinos
$\nu_1$ and $\nu_2$ are stable and relativistic, it is not difficult to derive
from (22) and (23) the probabilities that a solar $\nu_e$ with energy 
$E \cong |\vectorp| \equiv p$  will not 
change into $\nu_{\mu}$ on its way to the Earth, 
${P_{VO}(\nu_e \rightarrow \nu_e; t)}$, and  will 
transform into $\nu_{\mu}$ while traveling to the Earth,
${P_{VO}(\nu_e \rightarrow \nu_{\mu}; t)}$:
$$
{P_{VO}(\nu_e \rightarrow \nu_e; t)} =  |A_{ee}(t)|^2 
 = 1 - {1\over 2}
\sin^22\theta~\left[ 1 - \cos 2\pi {R(t_y)\over
L_v}\right], ~~~\eqno(24)
$$
$$
{P_{VO}(\nu_e \rightarrow \nu_{\mu}; t)} = |A_{\mu e}(t)|^2  
 = {1\over 2}
\sin^22\theta~\left[ 1 - \cos 2\pi {R(t_y)\over
L_v}\right], ~~~~~~~~\eqno(25)
$$
where $\Delta m^2 = m^2_{2} - m^2_{1}$,
$$
{L_v} = 4\pi {E}/\Delta m^2~~~\eqno(26)
$$ 
is the oscillation length in vacuum, 
$$
{R(t_y) = R_0~[ 1 - \epsilon\cos (2\pi t_y/T)}]~,~~~\eqno(27)
$$ 
is the Sun--Earth distance at time $t_y$ of the year
(T = 365 days), ${R_0} = 1.4966\times 10^8$ km and $\epsilon =
0.0167$ being the mean Sun--Earth distance and the ellipticity of the
Earth orbit around the Sun. In deriving (24) and (25) we have used
the equalities 
$$
E_2 - E_1 \cong p + \Delta m^2/(2p)\quad\mbox{and}\quad t \cong R(t_y)
$$
valid for relativistic neutrinos $\nu_{1,2}$. 
The quantities $\Delta m^2$ and $\sin^22\theta$ are typically 
considered and treated as free parameters to be determined by the
analysis of the solar neutrino data. 

It should be clear from the above discussion that the neutrino oscillations,
if they exist, would be a purely quantum mechanical phenomenon. 
The requirements of coherence between the states $|\nu_1\rangle$ and $|\nu_2\rangle$ in the
superposition  (18a) representing the $\nu_e$ at the production point, 
and that the coherence be maintained 
during the evolution of the neutrino system
up to the moment of neutrino detection, are crucial for the neutrino
oscillations to occur. The subtleties and the implications 
of the coherence condition for neutrino oscillations continue to be
discussed (see, e.g., \cite{NMP2,QMNO1,QMNO2} and the articles quoted
therein).

As it follows from (25), the $\nu_e \rightarrow \nu_{\mu}$
transition probability
${P_{VO}(\nu_e \rightarrow \nu_{\mu}; t)}$,
depends on two factors:
on $(1 - \cos 2\pi {R(t_y)/L_v})$, which exhibits oscillatory dependence on
the distance traveled by the neutrinos and on the neutrino energy (hence
the name ``neutrino oscillations''), and on  $\sin^22\theta$ which determines
the amplitude of the oscillations. In order for the 
$\nu_e \rightarrow \nu_{\mu}$
oscillation probability to be large, 
${P_{VO}(\nu_e \rightarrow \nu_{\mu}; t)} \cong 1$, 
two conditions have to be fulfilled: the neutrino mixing 
in vacuum must be large, $\sin^22\theta \cong 1$, and the oscillation length
in vacuum $L_v$ has to be of the order of or smaller than the 
distance traveled by the neutrinos, $R$: ${L_v \la 2\pi R}$. 
If the second condition
is not satisfied, i.e., if ${L_v \gg 2\pi R}$, the oscillations do not have 
enough time to develop on the way to the neutrino detector as the source-detector
distance $R$ (in our case the Sun--Earth distance) is too  short, and 
one has ${P_{VO}(\nu_e \rightarrow \nu_{\mu}; t)} \cong 0$.
 
Let us note that, in general, a given experiment searching for  
neutrino oscillations, is specified, in particular, by the average 
energy of the neutrinos being studied, $\bar{E}$,
and by the distance traveled by the neutrinos to the neutrino detector. 
The requirement ${L_v \la 2\pi R}$ determines the 
minimal value of the parameter $\Delta m^2$
to which the experiment is sensitive (figure of merit of the experiment):
${\rm min}(\Delta m^2) \sim 2\bar{E}/R$. 
Because of the interference nature of the neutrino oscillations,
the neutrino oscillation experiments can probe, in general, 
rather small values of
$\Delta m^2$ (see, e.g., \cite{NMP1,NMP2}).
In addition, due to the large distance 
between the Sun and Earth and the relatively low energies 
of the solar neutrinos, $\bar{E} \sim 1~{\rm MeV}$, 
the experiments with solar neutrinos have a remarkable 
sensitivity to the parameter $\Delta m^2$, namely, they can probe values as 
small as $10^{-11}~{\rm eV^2}$: 
$\Delta m^2 \ga 10^{-11}~{\rm eV^2}$.
 
To summarize the above discussion,  
if (18a) is realized and 
$\Delta m^2 \ga 10^{-11}~{\rm eV^2}$ 
the solar $\nu_e$ can take part in vacuum oscillations 
on the way to the Earth. In this case  
the flavour content of the electron neutrino
state vector will change periodically between the 
Sun and the Earth due to
the different time evolution of the vector's massive neutrino
components. The amplitude of these oscillations is determined
by the value of $\sin^22\theta$. 
If $\sin^22\theta$ is sufficiently large,  
the neutrinos that are being detected in the solar
neutrino detectors on Earth will be in states representing,
in general, certain superpositions of the states 
of\,\footnote{Obviously, if $\nu_e$ mixes with $\nu_{\mu}$ and/or
$\nu_{\tau}$ and/or $\nu_s$, these states will be superpositions
of the states of $\nu_{\mu}$ and/or
$\nu_{\tau}$ and/or $\nu_s$.} $\nu_e$ and $\nu_{\mu}$. 
As the muon (and tau and sterile) neutrinos 
interact much weaker with matter than electron 
neutrinos, the measured signals in the solar
neutrino detectors should be depleted with respect to the expected ones. 
This would explain the solar neutrino problem.

Detailed analyses of the solar neutrino data in terms of the    
hypothesis of two--neutrino 
vacuum oscillations of solar neutrinos
have been performed in the period after 1991, e.g.,
in \cite{KP1,KP11,KP3,SPnu96,SPHeid97}. 
It was found that the two--neutrino oscillations
involving the $\nu_e$ and an active neutrino,
$\nu_e\leftrightarrow\nu_{\mu (\tau)}$, provide 
a good quality description ($\chi^2$--fit) of the solar 
neutrino data for values of the two vacuum 
oscillation parameters belonging approximately to the region
(see, e.g., \cite{SPnu96}):
$$
5.0\times10^{-11} {\rm eV}^2\la \Delta m^2 \la 
10^{-10} {\rm eV}^2,~~\eqno(28a)
$$
$$
0.65 \la \sin^22\theta \leq 1.0.~~~\eqno(28b)
$$ 
At the same time, as it was shown in \cite{KP11,KP3},
the oscillations into sterile neutrino $\nu_s$,
$\nu_e\leftrightarrow\nu_s$, give a poor fit of the solar neutrino 
data and are thus strongly disfavored by the data as a 
possible solution of the solar neutrino problem. 

The probability of solar $\nu_e$ survival,
$P_{VO}(\nu_e \rightarrow \nu_{e}; R_0)$, in which
$t \cong R(t_y)$ is replaced with the average Sun-Earth distance 
$R_0$, and the probability $P_{VO}(\nu_e \rightarrow \nu_{e}; t)$
averaged over the period of one year,\footnote{The one-year 
averaged probability has to be used in the analyses
of data taken over periods of k years, k = 1,2,3,..., as are the data
(1) - (4) provided by the Cl-Ar, Ga-Ge and Kamiokande 
experiments.}
are shown for $\Delta m^2 = 6.3\times 10^{-11} {\rm eV}^2$ and 
$\sin^22\theta = 0.8$ as functions of the solar neutrino energy
$E$ in the upper and lower frames of Fig. 5, respectively (taken 
from \cite{KP3}).

\begin{figure}[t]
\begin{center}
\epsfig{file=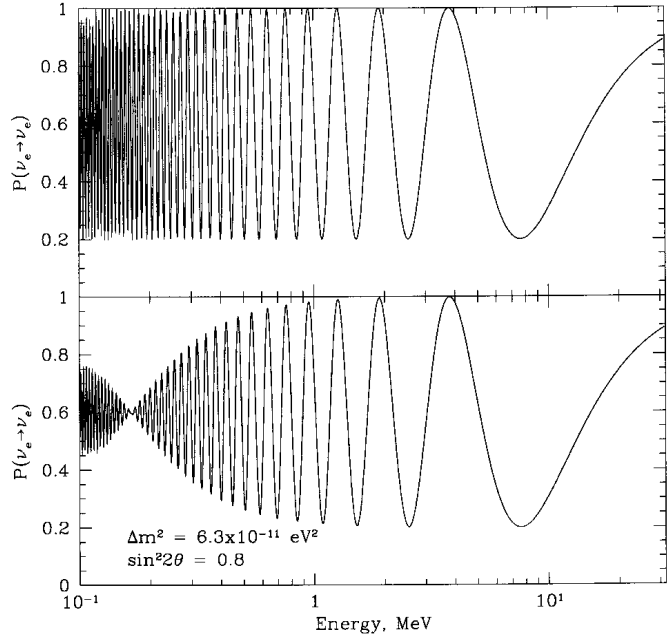,width=9cm}
\end{center}
\caption{The vacuum oscillation probability
$P_{VO}(\nu_e \rightarrow \nu_{e}; t)$, Eq. (24), for the mean
distance between the Sun and the Earth,
$t = R_0$ (upper frame), and the probability 
$P_{VO}(\nu_e \rightarrow \nu_{e}; t)$
averaged over a period of 1 year (lower frame), as functions of the neutrino
energy $E$ for $\Delta m^2 = 10^{-10}~{\rm eV^2}~$ and $\sin^22\theta =
0.8$ (from [57]
). For further details see the text.}
\end{figure}

Although in the analyses \cite{KP3,SPnu96} leading to the above results 
the predictions of the solar model \cite{BP95} 
with heavy element diffusion  
for the fluxes of the pp, pep, $^{7}$Be, $^{8}$B and CNO 
neutrinos were used, it was also verified \cite{KP3,BerRos,SPnu96}
that the results so obtained (i.e., the existence
of the vacuum $\nu_e\leftrightarrow\nu_{\mu (\tau)}$ 
oscillation solution) are stable with
respect to variations of the values of 
the total $^{8}$B and $^{7}$Be neutrino fluxes
within wide intervals in which the predictions of 
all contemporary solar models
lie.\footnote{Actually, in the analysis performed in 
\cite{KP3} the $^{8}$B neutrino flux $\Phi_{{\rm B}}$ was
treated as a free parameter, while the $^{7}$Be neutrino flux  
$\Phi_{{\rm Be}}$ was assumed to take values in the interval
$(0.7 - 1.3)\Phi^{\rm BP95}_{{\rm Be}}$, where $\Phi^{\rm BP95}_{{\rm B}}$ is
the flux in the model \cite{BP95} (see Table \ref{tab:effluents1}).}

\begin{figure}[t]
\begin{center}
\epsfig{file=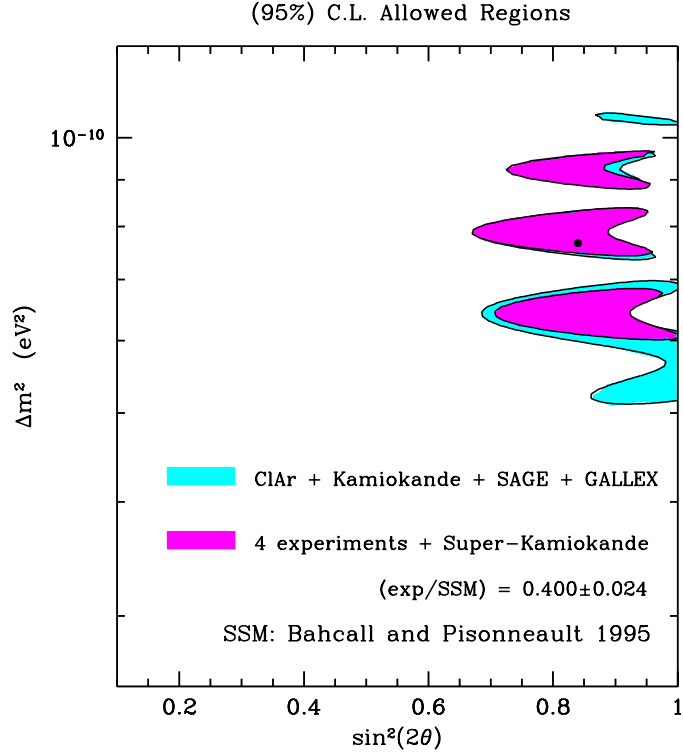,width=9cm}
\end{center}
\caption{Regions of values of the parameters 
$\Delta m^2$ and $\sin^22\theta$ (shown in black) for which the solar
neutrino data can be described (at 95\% C.L.) in terms of vacuum
$\nu_{e} \leftrightarrow \nu_{\mu (\tau)}$ oscillations of the solar $\nu_e$
(from [72]
). For further details see the text.}
\end{figure}

The results of a recent $\chi^2-$analysis of the solar neutrino data
in terms of the hypothesis of 
two-neutrino $\nu_e\leftrightarrow\nu_{\mu (\tau)}$ oscillations
of the solar neutrinos \cite{PK97}, are shown graphically in Fig. 6. 
The analysis was based
on the predictions of the solar model of Bahcall and Pinsonneault
\cite{BP95} with heavy element diffusion. 
The regions in the $\Delta m^2 - \sin^22\theta$ plane 
colored in black correspond to values of the two parameters for which one 
obtains (at 95\% C.L.) a description of the data, in other words,
a solution of the solar neutrino 
problem.\footnote{In this analysis the experimental results (1), (2), (4) and (6)
and the data from the GALLEX experiment obtained in 51 runs of measurements,
$\bar {\rm R}_{\rm GALLEX}({\rm Ge}) = 
(69.7~\pm 6.7^{+~3.9}_{-~4.5}) \hskip 0.2cm {\rm SNU}$,
were used. The solution regions of values of 
$\Delta m^2$ and $\sin^22\theta$ do not change significantly
if one uses the most recent GALLEX data, (4).}

Let us note that for the values of 
$\Delta m^2$ from the interval (28a), the oscillation length
in vacuum for the solar neutrinos with energy $E \sim 1~{\rm MeV}$ is of the
order of the Sun-Earth distance:
${L_v }\sim (2.5 - 5.0) \times 10^{7} ~{\rm km}$. At the same time
${L_v}$ is much bigger than the solar radius: ${\rm L_v \gg R_{\odot}}$.

\subsection{Matter-Enhanced Transitions}

Let us consider next that the possible 
effects of the solar matter on the 
oscillations of solar neutrinos 
assuming that  (18a) and (18b) hold true 
and supposing first that $|\nu_x\rangle \equiv |\nu_{\mu}\rangle$.

The presence of matter can drastically change the
pattern of neutrino oscillations: neutrinos can interact with the 
particles forming the matter.
Accordingly, the Hamiltonian of the neutrino system in matter 
differs from the Hamiltonian
of the neutrino system in vacuum $H_0$,
$$H_m = H_0 + H_{int}\,,~~\eqno(29)$$
where $H_{int}$ describes the interaction of the flavour 
neutrinos with the particles of matter.
When, e.g., electron neutrinos 
propagate in matter,
they can scatter (due to the $H_{int}$) on 
the particles present in matter: 
on the electrons (e$^{-}$), protons ($p$) and neutrons ($n$). 
The incoherent elastic 
and the quasi-elastic scattering, in which the 
states of the initial particles 
change in the scattering process  
(destroying the coherence between the neutrino states), 
are not of interest for our 
discussion for one simple reason - they have a 
negligible effect on the solar neutrino
propagation in the Sun
\footnote{These processes are important, however, for the supernova neutrinos
(see, e.g., \cite{NMP5,NMP6}).}~:
even in the center of the Sun, where
the density of matter is relatively high ($\sim 150~{\rm g/cm^3}$),
an $\nu_e$ with energy of 1 MeV has a mean free path with respect 
to the indicated scattering processes, which exceeds $10^{10}~{\rm km}$
(recall that the solar radius 
is much smaller: ${\rm R_{\odot}} = 6.96\times 10^{5}~{\rm km}$).
The oscillating $\nu_e$ and $\nu_{\mu}$ can scatter also elastically
in the forward direction on the e$^{-}$, $p$ and $n$, with
the momenta and the spin states of the 
particles participating in the elastic 
scattering reaction remaining unchanged.
In such a process the coherence of the neutrino states 
is being preserved and
the oscillations between the flavour neutrinos 
can continue in spite of, and actually, in parallel to, the scattering.

The $\nu_e$ and $\nu_{\mu}$ coherent elastic scattering in 
the forward direction on the particles of 
matter generates nontrivial indices of 
refraction of the $\nu_e$ and $\nu_{\mu}$ in matter \cite{MSW1}: 
$\kappa (\nu_e) \neq 1$, $\kappa (\nu_{\mu}) \neq 1$. 
Most importantly,
the index of refraction of the $\nu_e$ thus generated does not coincide
with the index of refraction of the $\nu_{\mu}$:
$\kappa (\nu_e) \neq \kappa (\nu_{\mu})$. The difference between the
two indices of refraction is determined essentially by the
difference of the real parts of the 
forward $\nu_e - e^{-}$ and $\nu_{\mu} - e^{-}$ elastic 
scattering amplitudes \cite{MSW1}, 
${\rm Re~[F_{\nu_e-e^{-}}(0)] - Re~[F_{\nu_{\mu}-e^{-}}(0)]}$: 
due to the flavour symmetry of the
neutrino -- quark (neutrino -- nucleon) 
neutral current interaction, 
the forward $\nu_e - p,n$ and $\nu_{\mu} - p,n$ elastic 
scattering amplitudes are equal and therefore do not contribute 
to the difference of interest.%
\footnote{We standardly assume that the weak interaction of the flavour
neutrinos $\nu_e$, $\nu_{\mu}$ and $\nu_{\tau}$ and antineutrinos
$\bar{\nu}_e$, $\bar{\nu}_{\mu}$ and $\bar{\nu}_{\tau}$ is described by 
the standard (Glashow-Salam-Weinberg) theory of 
electroweak interaction and 
that the generation of nonzero neutrino 
masses and lepton mixing leading to
(18a) and (18b) does not produce new couplings 
which can change substantially the neutrino weak 
interaction, as required by the 
existing experimental limits on such new couplings
(for an alternative possibility see, e.g., \cite{FCNC1}).
Let us add that the imaginary parts of the 
forward scattering amplitudes
(responsible, in particular, for decoherence effects)
are proportional to the corresponding total scattering 
cross-sections and in the case of interest are negligible
in comparison with the real parts.}
The real parts of the amplitudes 
${\rm F_{\nu_e-e^{-}}(0)}$ and ${\rm F_{\nu_{\mu}-e^{-}}(0)}$ 
can be calculated in the standard theory. One finds 
the following result \cite{MSW1,Barger80,Langa83} (see also \cite{JLiu91})
for the difference of the indices of refraction of $\nu_e$ and $\nu_{\mu}$:
$$
\kappa (\nu_e) - \kappa (\nu_{\mu}) =  {2\pi \over{p^2}}~
\left( {\rm Re~[F_{\nu_e-e^{-}}(0)] - Re~[F_{\nu_{\mu}-e^{-}}(0)]} \right) =
 - ~{1\over{p}}\sqrt{2}G_{F}N_{e}~, ~\eqno(30)
$$
where $G_{F}$ is the Fermi constant and  $N_{e}$ is the
electron number density in matter. Let us note that 
the forward scattering amplitudes for the antineutrinos
${\rm F_{\bar{\nu}_e-e^{-}}(0)}$ and 
${\rm F_{\bar{\nu}_{\mu}-e^{-}}(0)}$ coincide in absolute value with
the amplitudes ${\rm F_{\nu_e-e^{-}}(0)}$ and 
${\rm F_{\nu_{\mu}-e^{-}}(0)}$
but have opposite sign and therefore one has
$$
\kappa (\bar{\nu}_e) - \kappa (\bar{\nu}_{\mu}) = 
+~ {1\over{p}}\sqrt{2}\,G_{F}\,N_{e}~. ~\eqno(31)
$$

Knowing the expression for the difference of the indices of refraction of
$\nu_e$ and $\nu_{\mu}$ in matter, it is not difficult to write the system of 
evolution equations which describes the  $\nu_e \leftrightarrow \nu_{\mu}$
oscillations in matter \cite{MSW1,Barger80,Langa83}:
$$
i\frac{d}{dt}\left(\begin{array}{c}A_{e}(t,t_0)\\A_{\mu}(t,t_0)\end{array}
\right)=\left(\begin{array}{cc}
-\epsilon(t)  & \epsilon'\\
 \epsilon' & \epsilon(t)
\end{array}\right)
\left(\begin{array}{c}A_{e}(t,t_0)\\A_{\mu}(t,t_0)\end{array}\right)
~~\eqno(32)
$$
where $A_{e}(t,t_0)$ ($A_{\mu}(t,t_0)$) is the amplitude
of the probability to find neutrino $\nu_{e}$ ($\nu_{\mu}$) at time $t$
of the evolution of the neutrino system if at time $t_0$ the neutrino
$\nu_{e}$ or $\nu_{\mu}$ (or a state representing a 
linear combination of the states describing the
two neutrinos) has been produced, $t \geq t_0$. Furthermore
$\epsilon (t)$ and $\epsilon'$ are real functions of 
the neutrino energy $E \cong p$, of 
$\Delta m^2$, of the mixing angle in vacuum $\theta$ and of
the electron number density in the point of the neutrino trajectory in matter
reached at time $t$, $N_{e}(t)$,
$$\epsilon (t) = {1\over {2}}~[ {~\Delta m^2 \over{2E}}\cos2\theta -
\sqrt{2} G_{F}N_{e}(t)],
~~\epsilon' = {\Delta m^2 \over{4E}}\sin 2\theta. ~~~\eqno(33)$$
The term $\sqrt{2} G_{F}N_{e}(t)$ in the parameter 
$\epsilon (t)$ accounts for the
effects of matter on the neutrino oscillations.

Let us note that the system of evolution equations describing the
oscillations of antineutrinos $\bar{\nu}_e \leftrightarrow \bar{\nu}_{\mu}$
in matter has exactly the same form except for the matter term
in $\epsilon(t)$ which, in accordance with (30) and (31), changes sign.

Due to the presence of the interaction term $H_{int}$ in the Hamiltonian
of the neutrino system in matter $H_{m}$, the eigenstates of the 
Hamiltonian of the neutrino system in vacuum,
$|\nu_1\rangle$ and $|\nu_2\rangle$, are not eigenstates of $H_{m}$. As a result of the
coherent scattering of $\nu_e$ and $\nu_{\mu}$ off the particles forming the matter  
transitions between the states $|\nu_1\rangle$ and $|\nu_2\rangle$ become possible
in matter:
$$ \langle\nu_2|~H_{int}~|\nu_1\rangle~ \neq ~0.~~~\eqno(34)$$
Consider first the case of $\nu_e \leftrightarrow \nu_{\mu}$ 
oscillations taking place in matter with 
electron number density which does not change along 
the neutrino trajectory:
$N_{e}(t) = N_{e} = ~const.$ 
It proves convenient to find the states
$|\nu^{m}_{1,2}\rangle$, which diagonalize the evolution matrix
in the right-hand side of the system (32), or equivalently, the 
Hamiltonian of the neutrino system in matter. The relations between
the matter-eigenstates $|\nu^{m}_{1,2}\rangle$ and the flavour-eigenstates
$|\nu_{e,\mu}\rangle$ have the same form as the relations (18a) and (18b) 
between the vacuum mass-eigenstates $|\nu_{1,2}\rangle$ and $|\nu_{e,\mu}\rangle$:
$$
|\nu_{e}\rangle~ = ~~~|\nu^{m}_{1}\rangle\cos\theta_{m} +
                   |\nu^{m}_{2}\rangle\sin\theta_{m},~~~\eqno(35a)
$$
$$
|\nu_{\mu}\rangle~~ = - |\nu^{m}_{1}\rangle\sin\theta_{m} ~+
                   |\nu^{m}_{2}\rangle\cos\theta_{m}~.~~~\eqno(35b)
$$
Here $\theta_{m}$ is the neutrino mixing angle in matter \cite{MSW1},
$$\sin2\theta_{m} = {\epsilon'\over{\sqrt{\epsilon^{2} + \epsilon'^{2}}}}
= {\tan2\theta \over
{\sqrt{(1 - {N_e \over {N_e^{res}}})^2 + \tan^22\theta}}},~~~\eqno(36)
$$
$$\cos2\theta_{m} = {\epsilon\over{\sqrt{\epsilon^{2} + \epsilon'^{2}}}} = 
{{1 - N_e/N_e^{res}}\over
{\sqrt{(1 - {N_e \over {N_e^{res}}})^2 + \tan^22\theta}}},~~~\eqno(37)
$$
where the quantity 
$$N_e^{res}  = 
{{\Delta m^2 \cos2\theta} \over { 2E \sqrt{2}  G_F}}~~~\eqno(38)$$
is called ``resonance density'' \cite{Barger80}. 
The matter-eigenstates $|\nu^{m}_{1,2}\rangle$ (which are also called ``adiabatic'')
are eigenstates of the evolution matrix (Hamiltonian)
in (32), corresponding to the two eigenvalues,
$E^{m}_{1,2}$, whose
difference is given by
$$E^{m}_{2} - E^{m}_{1} = 2\sqrt{\epsilon^{2} +
\epsilon'^{2}} = 
{\Delta m^2 \over {2E}}~ 
\sqrt{(1 - {N_e \over {N_e^{res}}})^2\cos^22\theta + \sin^22\theta}.~~~\eqno(39)$$

It should be almost obvious from  (25) after 
comparing (18a), (18b) with (35a), (35b)
that the probability to find neutrino $\nu_{\mu}$ at time $t$ if neutrino
$\nu_e$ has been produced at time $t_0 = 0$ and it traversed a distance
$(t - t_0) = t \cong R_{m}$ in matter with constant electron number density
$N_e$, has the form \cite{MSW2}:  
$$P_{m}(\nu_e \rightarrow \nu_{\mu}; t) = |A_{\mu}(t)|^2  
 = {1\over 2}
\sin^22\theta_{m}~\left[ 1 - \cos 2\pi {R_{m}\over
 L_{m}}\right],~~~\eqno(40)
$$
where
$$ 2\pi {R_{m}\over L_{m}} \cong (E^{m}_{2} - E^{m}_{1})t,~~\eqno (41)$$
and $L_{m}$ is the oscillation length in matter:
$$L_{m} = {L_v \over {\sqrt{(1 - {N_e \over {N_e^{res}}})^2 \cos^22\theta + 
\sin^22\theta}}}\,.~\eqno(42)$$

Evidently, the amplitude of the $\nu_e \leftrightarrow \nu_{\mu}$ 
oscillations in matter is equal to $\sin^22\theta_{m}$. It follows from
(36) that, most remarkably, the dependence of 
$\sin^22\theta_{m}$ on $N_e$ has a resonance character 
\cite{MSW2}. Indeed, if in the case 
of interest the condition 
$$
\Delta m^2 \cos^22\theta > 0~~~\eqno(43)
$$ 
is fulfilled, for any finite value of $\sin^22\theta$
there exists a value of the electron number density 
equal to $N_e^{res}$,
such that when 
$$N_e = N_e^{res}~~\eqno(44)
$$ 
we have
$$
\sin^2 2\theta_{m} = 1\,.~~\eqno(45)
$$
Note that if $N_e = N_e^{res}$, we get $\sin^22\theta_{m} = 1$
even if the mixing angle in vacuum is small, i.e., if $\sin^22\theta \ll 1$.
This implies that the presence of matter can lead to a strong enhancement
of the oscillation probability $P_{m}(\nu_e \rightarrow \nu_{\mu}; t)$
even when the $\nu_e \leftrightarrow \nu_{\mu}$ oscillations
in vacuum are strongly suppressed due to a small value of  
$\sin^22\theta$ (hence, the name ``matter-enhanced neutrino oscillations'').  

The oscillation length at resonance is given by
$$
L^{res}_{m} = {L_v \over {\sin2\theta}}\,,~\eqno(46
$$
while the width in $N_{e}$ of the resonance (i.e., the ``distance'' in
$N_{e}$ between the points at which $\sin^22\theta_{m} = 1/2$) has the form 
$$
\Delta N_e^{res} = 2N_e^{res}\tan2\theta~.~\eqno(47)
$$
Thus, if the mixing angle in vacuum is small the resonance is narrow,
$\Delta N_e^{res} \ll N_e^{res}$,
and the oscillation length in matter at resonance is relatively large,
$L^{res}_{m} \gg L_v$. As it follows from  (39), the energy 
difference $E^{m}_{2} - E^{m}_{1}$ has a minimum at the resonance:
$$
(E^{m}_{2} - E^{m}_{1})^{res} =  {\rm min}~(E^{m}_{2} - E^{m}_{1}) 
=  {\Delta m^2 \over {2E}}~ \sin 2\theta.~~~\eqno(48)
$$

It is instructive to consider two limiting case. 
If $N_e \ll N_e^{res}$, as it follows from  (36) and (42),
$\theta_{m} \cong \theta$
($\sin2\theta_{m} \cong \sin2\theta$), ${\rm L_{m}} \cong {{\rm L_v}}$
and the neutrinos oscillate practically as in vacuum. In the opposite limit,
$N_e \gg N_e^{res},~N_e^{res}\tan^22\theta$, one finds from  (36) and (37) that
$\theta_{m} \cong \pi/2$ ($\sin2\theta_{m} \cong 0$, $\cos2\theta_{m} \cong -1$)
and the presence of matter suppresses the 
$\nu_e \leftrightarrow \nu_{\mu}$ oscillations (see  (40)).
In this case we get from  (35a) and (35b):
$$
|\nu_{e}\rangle~ \cong ~~~|\nu^{m}_{2}\rangle,~~~\eqno(49a)
$$
$$
|\nu_{\mu}\rangle~ = - |\nu^{m}_{1}\rangle,~~~\eqno(49b)
$$
i.e., if the electron number density exceeds considerably the
resonance density,  $\nu_e$ practically coincides with the heavier
of the two matter-eigenstate neutrinos $\nu^{m}_{2}$, while the 
$\nu_{\mu}$ coincides with the lighter one $\nu^{m}_{1}$.  

The analogs of  (36), (37), (39), (40) and (42) for 
oscillations of antineutrinos,
$\bar{\nu}_e \leftrightarrow \bar{\nu}_{\mu}$,
in matter with constant $N_e$ can formally be obtained 
by replacing $N_e$ with $(- N_e)$ in the indicated equations.
If condition (43) is fulfilled, we have $N_e^{res} > 0$ and the term
$(1 +  N_e/ N_e^{res})$ which appears, e.g., in the expression for the
mixing angle in matter $\bar{\theta}_{m}$ in the case of 
$\bar{\nu}_e \leftrightarrow \bar{\nu}_{\mu}$ oscillations,
can never be zero. Thus, a resonance enhancement of the
$\bar{\nu}_e \leftrightarrow \bar{\nu}_{\mu}$ oscillations
cannot take place. The matter, actually, can only suppress
the oscillations. 

It should be clear from this discussion that depending on the sign of the
product $\Delta m^2~\cos 2\theta$, the presence of 
matter can lead to resonance
enhancement either of the 
$\nu_e \leftrightarrow \nu_{\mu}$ or of the
$\bar{\nu}_e \leftrightarrow \bar{\nu}_{\mu}$ oscillations, but not 
of the both types of oscillations.  
This is a consequence of the fact 
\cite{LangP}
that the matter in the Sun or in the Earth we are 
interested in, is not charge-symmetric (it contains $e^{-},~p$ and $n$, 
but does not contain their antiparticles) and therefore the oscillations
in matter are neither  CP- nor CPT- invariant.%
\footnote{As it is not difficult to convince oneself,
the matter effects in the 
$\nu_e \leftrightarrow \nu_{\mu}$ 
($\bar{\nu}_e \leftrightarrow \bar{\nu}_{\mu}$) oscillations
will be invariant with respect
to the operation of time reversal if the 
$N_e$ distribution along the neutrino path is symmetric with respect
to this operation. The latter condition is fulfilled for the 
$N_e$ distribution along a path of a neutrino crossing the Earth
\cite{KPEarth}.}
In what follows we shall assume that
$\Delta m^2 > 0$ and $\cos 2\theta > 0$, so that  (43) is satisfied
and therefore only the $\nu_e \leftrightarrow \nu_{\mu}$ 
oscillations can be enhanced by the matter effects. 

Since the neutral current weak interaction of neutrinos in the standard
theory is flavour symmetric, the formulae and results we have 
obtained above and shall obtain in what follows are valid for the
case of $\nu_e - \nu_{\tau}$ mixing ((18a) and (18b)) and 
$\nu_e \leftrightarrow \nu_{\tau}$ 
oscillations in matter as well. In what concerns the possibility 
of mixing and oscillations between the $\nu_e$ and a sterile neutrino $\nu_s$,
$\nu_e \leftrightarrow \nu_{s}$, the relevant formulae can be obtained
from the formulae derived for the case of 
$\nu_e \leftrightarrow \nu_{\mu (\tau)}$ oscillations 
by \cite{LangP} replacing $N_e$ with $(N_e - 1/2N_n)$,
where $N_n$ is the number density of neutrons in matter.

The formalism we have developed above can be directly 
applied, for instance,
to the study of the matter effects in the 
$\nu_e \leftrightarrow \nu_{\mu (\tau)}$
($\nu_{\mu (\tau)} \leftrightarrow \nu_{e}$)
oscillations of the flavour neutrinos
which traverse the Earth mantle 
(but do not traverse the Earth core). The electron number density
changes little around the mean value of 
$\bar{N}_e \cong 2.3~cm^{-3}~N_{A}$,
$N_{A}$ being the Avogadro number,
along the trajectories of neutrinos which cross a 
substantial part of the Earth mantle and the  
$N_e = const.$ approximation is rather accurate. If, for example,
$\Delta m^2 = 10^{-3}~{\rm eV^2}$, $E = 1~{\rm GeV}$ and $\sin^22\theta \cong 0.5$,
we have: $N_e^{res} \cong 4.6~cm^{-3}~N_{A}$, 
$\sin^22\theta_{m} \cong 0.8$ and the oscillation length in matter,
$L_{m} \cong 3\times 10^{3}~{\rm km}$, is of the order of the 
depth of the Earth mantle,\footnote{The Earth 
radius is 6371 km; the Earth core, 
whose density ($N_e$) is larger approximately by a factor of 2.5 
than the density ($N_e$) in the mantle,
has a radius of 3486 km, so the Earth mantle depth is 2885 km.}
so that one can have $2\pi R_{m} \ga L_m$.
It is not difficult to obtain an expression for the 
$\nu_e \leftrightarrow \nu_{\mu(\tau)}$ oscillation 
probability in the case when the neutrinos traverse both the
Earth mantle and the core assuming 
${N_e}$ is constant, but has different values
in the two Earth density structures. 

It is not clear, however, what the above 
interesting results have to do with the problem 
of main interest for us, namely, 
accounting for the effects of solar matter in the
oscillations of solar neutrinos while they propagate 
from the central part to the surface of the Sun. 
The electron number density (the matter density) 
changes considerably along the neutrino path in the Sun:
it decreases monotonically from the value
of $\sim 100~{\rm cm^{-3}}~N_{A}$ ($\sim 150~{\rm g/cm^{3}}$)
in the center of the Sun to 0 at the surface of the Sun.
Actually, according to the contemporary solar models (see, e.g., 
\cite{JNB,BP95}), 
$N_e$ decreases approximately exponentially 
in the radial direction towards
the surface of the Sun:  
$$
N_e(t) = N_e(t_0)\exp{\left\{- {{t - t_0}\over{r_0}}\right\}},
~~~~~\eqno(50)
$$
where $(t - t_0) \cong d$ is the distance traveled
by the neutrino in the Sun, $N_e(t_0)$ is the electron number 
density in the point
of neutrino production in the Sun,
$r_0$ is the scale-height of the change of $N_e(t)$ 
and one has \cite{JNB} ~$r_0 \sim 0.1R_{\odot}$.

Obviously, if $N_e$ changes with $t$ 
(or equivalently with the distance) along 
the neutrino trajectory,
the matter-eigenstates, their energies, 
the mixing angle and the oscillation length in matter,
become, through their dependence on $N_e$,
also functions of $t$:
$|\nu^{m}_{1,2}\rangle = |\nu^{m}_{1,2}(t)\rangle$,
$E^{m}_{1,2} = E^{m}_{1,2}(t)$,
$\theta_{m} = \theta_{m}(t)$ and 
$L_m = L_{m}(t)$. 

It is not difficult to understand 
qualitatively the possible behaviour of the neutrino system when 
solar neutrinos propagate from the center to the surface of the Sun
if one realizes that one is dealing effectively with a two-level 
system whose Hamiltonian depends on time 
and admits``jumps'' from one level to the other (see  (32)).
Let us assume first for simplicity that 
the electron number density in the point 
of a solar $\nu_e$ production in the Sun 
is much bigger than the resonance density, 
$N_e(t_0) \gg N_e^{res}$,
and that the mixing angle in vacuum is small,
$\sin \theta \ll 1$. Actually, this is one of the cases relevant to 
the solar neutrinos. In this case we have 
$\theta_{m}(t_0) \cong \pi/2$ and the 
state of the electron neutrino in
the initial moment of the evolution of the 
system practically coincides with the heavier of the 
two matter-eigenstates:
$$
|\nu_{e}\rangle~ \cong ~~~|\nu^{m}_{2}(t_0)\rangle.~~~\eqno(51)
$$
Thus, at $t_0$ the neutrino system is in a state
corresponding to the ``level'' with energy 
$E^{m}_{2}(t_0)$. When neutrinos propagate  
to the surface of the Sun they cross a layer of matter 
in which $N_e = N_e^{res}$: in this layer the
difference between the energies of the two ``levels''
$(E^{m}_{2}(t) - E^{m}_{1}(t))$ has a minimal value  
on the neutrino trajectory ( (39) and (40)).
Correspondingly, the evolution of the neutrino system 
can proceed basically in two ways.
First, the system can stay on the ``level'' with energy
$E^{m}_{2}(t)$, i.e., can continue to be in the state
$|\nu^{m}_{2}(t)\rangle$ up to the final moment $t_{s}$, 
when the neutrino reaches the surface of the Sun.
At the surface of the Sun 
$N_e(t_s) = 0$ and therefore 
$\theta_{m}(t_s) = \theta$,
$|\nu^{m}_{1,2}(t_{s})\rangle \equiv |\nu_{1,2}\rangle$ and 
$E^{m}_{1,2}(t_s) = E_{1,2}$. 
Thus, in this case the state describing 
the neutrino system at $t_0$ will evolve continuously 
into the state $|\nu_{2}\rangle$ at the surface of the Sun.
Using  (18a) and (18b), it is trivial 
to obtain now the probabilities to find
respectively neutrino $\nu_{e}$ and neutrino $\nu_{\mu}$ at the surface 
of the Sun (given the fact that $\nu_e$ has been 
produced in the initial point of the neutrino trajectory):
$$P(\nu_e \rightarrow \nu_{e};t_s, t_0) \equiv
|A_{e}(t_s,t_0)|^2 \cong |\langle\nu_{e} |\nu_{2}\rangle|^2 = \sin^2\theta, ~~\eqno(52a)
$$
$$
P(\nu_e \rightarrow \nu_{\mu};t_s, t_0) \equiv
|A_{\mu}(t_s,t_0)|^2 \cong |\langle\nu_{\mu} |\nu_{2}\rangle|^2 = \cos^2\theta.~~\eqno(52b)
$$
It is clear that under the assumptions made (i.e., $\sin^2\theta \ll 1$), 
a practically total $\nu_e - \nu_{\mu}$ conversion is possible
in the case under study. This type of evolution of 
the neutrino system as well as the 
$\nu_e \rightarrow \nu_{\mu}$ transitions taking 
place during the evolution, are
called \cite{MSW2} ``adiabatic''. They are characterized by the fact that the 
probability of the ``jump'' from the upper ``level'' (having energy 
$E^{m}_{2}(t)$)
to the lower ``level'' (with energy $E^{m}_{1}(t)$),
$P'$, or equivalently the probability   
of the $\nu^{m}_{2}(t_{0}) \rightarrow \nu^{m}_{1}(t_{s})$
transition, $P' \equiv P'(\nu^{m}_{2}(t_{0}) \rightarrow \nu^{m}_{1}(t_{s}))$,
on the whole neutrino trajectory
is negligible:
$$
P' \equiv P'(\nu^{m}_{2}(t_{0}) \rightarrow \nu^{m}_{1}(t_{s})) \cong 0~:
~adiabatic ~transitions.~~\eqno(53)
$$
    
The second possibility is realized if in the resonance region, where
the two ``levels'' approach each other most (the
difference between the energies of the two ``levels'' 
$(E^{m}_{2}(t) - E^{m}_{1}(t))$ has a minimal value),
the system ``jumps'' from the upper ``level'' 
to the lower ``level'' and after that continues to be in the state
$|\nu^{m}_{1}(t)\rangle$ until the neutrino reaches the surface of the Sun.
Evidently, now we have
$P' \equiv P'(\nu^{m}_{2}(t_{0}) \rightarrow \nu^{m}_{1}(t_{s})) \cong 1$.
In this case the neutrino system ends up in the state
$|\nu^{m}_{1}(t_{s})\rangle \equiv |\nu_{1}\rangle$
at the surface of the Sun and the probabilities to find the neutrinos
$\nu_e$ and $\nu_{\mu}$ at the surface of the Sun are given by
$$
P(\nu_e \rightarrow \nu_{e};t_s, t_0) \equiv
|A_{e}(t_s,t_0)|^2 \cong |\langle\nu_{e} |\nu_{1}\rangle|^2 = \cos^2\theta, ~~\eqno(54a)
$$
$$
P(\nu_e \rightarrow \nu_{\mu};t_s, t_0) \equiv
|A_{\mu}(t_s,t_0)|^2 \cong |\langle\nu_{\mu} |\nu_{1}\rangle|^2 = \sin^2\theta.~~\eqno(54b)
$$
Obviously, if $\sin^2\theta \ll 1$, practically no
transitions of the solar $\nu_e$ into $\nu_{\mu}$ will occur.
The considered regime of evolution of 
the neutrino system and the corresponding 
$\nu_e \rightarrow \nu_{\mu}$ transitions are usually referred to as
``extremely nonadiabatic''. 

Clearly, the value of the ``jump'' probability
$P'$ plays a crucial role in the  
the $\nu_e \rightarrow \nu_{\mu}$ transitions: 
it fixes the type of the transition and determines to large extent  
$\nu_e \rightarrow \nu_{\mu}$ transition probability. 
We have considered above two limiting cases: 
$P'\cong 0$ and $P'\cong 1$. Obviously, there exists a whole spectrum of
possibilities since $P'$ can have any value from 0 to 1. In general,
the transitions are called ``nonadiabatic'' 
if $P'$ is non-negligible (see further).

Numerical studies have shown \cite{MSW2} that solar neutrinos can undergo
both adiabatic and nonadiabatic $\nu_e \rightarrow \nu_{\mu}$ transitions
in the Sun and the matter effects can be substantial in the solar neutrino
oscillations for a remarkably wide range of values of 
the two parameters $\Delta m^2$ and $\sin^22\theta$, namely for
$$
10^{-7} {\rm eV}^2\la \Delta m^2 \la 
10^{-4} {\rm eV}^2,~~\eqno(55a)
$$
$$
10^{-4} \la \sin^22\theta \leq 1.0\,.~~~\eqno(55b)
$$ 

It would be preferable to make more quantitative 
the preceding analysis. We will obtain first the adiabaticity
condition \cite{MSW2,Messiah86}.

Using the   (35a) and (35b) we can express the probability amplitudes
$A_{e}(t,t_0)$  and $A_{\mu}(t,t_0)$
in terms of the probability amplitudes
$A_{1}(t,t_0)$ and $A_{2}(t,t_0)$ to find the neutrino
system in the states $|\nu^{m}_{1}(t)\rangle$ and $|\nu^{m}_{2}(t))\rangle$, 
respectively, at time $t$: 
$$
A_{e}(t,t_0)~ = ~~~A_{1}(t,t_0)\cos\theta_{m}(t) +
                   A_{2}(t,t_0)\sin\theta_{m}(t),~~~\eqno(56a)
$$
$$
A_{\mu}(t,t_0)~ = - A_{1}(t,t_0)\sin\theta_{m}(t) +
                     A_{2}(t,t_0)\cos\theta_{m}(t).~~~\eqno(56b)
$$
Substituting  (56a) and (56b) in  (32) 
we obtain the system of evolution equations 
for the probability amplitudes  
$A_{1}(t,t_0)$ and $A_{2}(t,t_0)$:
\vskip 0.2truecm
$$
i\frac{d}{dt}\left(\begin{array}{c}A_{1}(t,t_0)\\A_{2}(t,t_0)\end{array}
\right)=\left(\begin{array}{cc}
 E_{1}^{m}(t)  & -i\dot{\theta}_{m}(t) \\
 i\dot{\theta}_{m}(t) & E_{2}^{m}(t) 
\end{array}\right)
\left(\begin{array}{c}A_{1}(t,t_0)\\A_{2}(t,t_0)\end{array}\right)\,.
~~\eqno(57)
$$
Here $\dot{\theta}_{m}(t) \equiv {d\over{dt}} \theta_{m}(t)$.
It follows from the preceding discussion that the solar neutrino
transitions in the Sun will be adiabatic (nonadiabatic) if the nondiagonal
term in the evolution matrix in the right-hand side of  (57),
which is responsible for the  
$\nu^{m}_{2}(t_{0}) \rightarrow \nu^{m}_{1}(t_{s})$
transitions, is sufficiently small (is non-negligible). The corresponding 
conditions can be written as
$$
4n(t) \gg 1,~~~~~ adiabatic~~transitions,~~~~~~~~\eqno(58a)
$$
$$
4n(t) \la 1,~~~~~ nonadiabatic~~transitions,~~\eqno(58b)
$$
where the adiabaticity function $4n(t)$ is given by
$$
4n(t) \equiv {{E^{m}_{2}(t) - E^{m}_{1}(t)}\over{2|\dot{\theta}_{m}(t)|}} =
\sqrt{2} G_F {{(N^{res}_e)^2}\over{|\dot{N}_e(t)|}} \tan^22\theta
\left( 1 + tan^{-2}2\theta_{m}(t) \right)^{\frac {3}{2}}~.~~\eqno(59)
$$
In  (59) $\dot{N}_e(t) \equiv {d\over{dt}}N_e(t)$
and we have used  (36), (37) and (39) to derive it. Expression (59)
for $4n(t)$ implies that the solar neutrino transitions in the Sun 
will be adiabatic if the electron number density changes sufficiently slowly
along the neutrino trajectory; if the change of $N_e(t)$ is relatively
fast, the transitions would be nonadiabatic.

In order for the solar neutrino transitions to be, e.g., adiabatic, 
condition (58a) has to be fulfilled in any point of neutrino trajectory
in the Sun. However, it is not difficult to convince oneself 
using  (36), (37), (50) and (59) that if the solar neutrinos cross a layer
with resonance density $N_e^{res}$ on their way to the surface of the Sun,
condition (58a) will hold if it holds at the resonance point, 
i.e., for the parameter
$$
\begin{array}{rcl}
4n_0 \equiv 4n(t = t_{res}) &=&
\displaystyle
\sqrt{2} G_F {{(N^{res}_e)^2}\over{|\dot{N}_e(t = t_{res})|}} \tan^22\theta
\nonumber\vspace{2mm}\\
&=&
\displaystyle
 r_0~{\Delta m^2 \over{2E}}~{\sin^22\theta \over{\cos2\theta}} = 
\pi ~ {\Delta r^{res} \over{L_{m}^{res}}}\, ,
\end{array}~~\eqno(60)
$$
where $t_{res}$ is the time at which the resonance layer is
crossed by the neutrinos, $t_0 < t_{res} < t_s$, 
$\Delta r^{res} = 2(N^{res}_e/|\dot{N}_e(t = t_{res})|)\tan 2\theta
\cong 2r_0~\tan 2\theta$ is the spatial width of the resonance and we have used
(38) and (46). Thus, the value of the adiabaticity parameter
$4n_0$ determines the type of the solar neutrino transitions.
It follows from  (60), in particular, that the transitions will be adiabatic
if the width of the resonance is bigger than the oscillation length
at resonance.

Actually, the system of evolution equations (32) can be solved exactly
in the case when $N_e$ changes exponentially,  (50), along the neutrino path in
the Sun \cite{SP200,EXP}. On the basis of the exact solution, 
which is expressed in terms of confluent hypergeometric
functions \cite{BE}, it was possible to derive a complete, simple
and very accurate analytic description of the matter-enhanced
transitions of solar neutrinos in the Sun 
\cite{SP200}, \cite{SP191}--\cite{SPJR89} (for a review see \cite{SP90}).
The probability that a $\nu_e$ having
momentum p (or energy ${\rm E \cong p}$) and produced at time t$_0$ in
the central part of the Sun will not transform into
$\nu_{\mu (\tau)}$ on its way to the surface
of the Sun (reached at time t$_s$) is given by
$$
P_{\odot}(\nu_e \rightarrow \nu_{e};t_s, t_0) = 
\bar{P}_{\odot}(\nu_e \rightarrow \nu_{e};t_s, t_0) +~Oscillating~terms. 
~~\eqno(61)
$$
Here
$$
\bar{P}_{\odot}(\nu_e \rightarrow \nu_{e};t_s, t_0) \equiv 
\bar{P}_{\odot} = {1\over 2} + \left({1\over2} -
{\rm P}^{'}\right)\cos2\theta_m(t_0) \cos2\theta~\eqno(62)
$$
is the average probability, where
$$P^{'} = {{\exp\left[-2\pi{\rm r}_0{\Delta m^2\over{2E}}\sin^2\theta
\right] - \exp\left[-2\pi{\rm r}_0{\Delta m^2\over{2E}}\right]}\over
{1 - \exp\left[-2\pi{\rm r}_0{\Delta m^2\over{2E}}\right]}}
~~~~~~~~~~~~~~~~~~~~~~~~~~~~~~~
$$
$$
= ~{{\exp\left[-2\pi n_{0}(1 - \tan^2\theta)\right]
          - \exp\left[-2\pi n_{0}(\tan^{-2}\theta - \tan^2\theta)\right]}
    \over {1 - \exp\left[-2\pi n_{0}(\tan^{-2}\theta - \tan^2\theta)\right]}}~
,~~\eqno(63)
$$
is \cite{SP200,SP90} the ``jump'' probability 
for exponentially varying electron number density%
\footnote{An expression for the ``jump'' probability
corresponding to the case of density ($N_e(t)$) varying linearly
along the neutrino path was derived a long time ago by Landau and
Zener \cite{LZ}. An analytic description of the solar neutrino transitions
based on the linear approximation for the change of $N_e$ in the Sun
and on the Landau-Zener result was proposed in \cite{LINEAR}.
The drawbacks of this description, which is less accurate \cite{KP88}
than the description based on the results obtained in the exponential
density approximation, were discussed, e.g., in \cite{SP191,KP88,SP90}.} 
~$N_e$, and $\theta_{m}(t_0)$ is the neutrino mixing angle in matter in the
point of $\nu_e$ production. 

We will not give the explicit analytic expressions for the
oscillating terms in the probability
$P_{\odot}(\nu_e \rightarrow \nu_{e};t_s, t_0)$, although they
have been derived in the exponential density
approximation for the $N_e$ as well \cite{SP88osc} (see also \cite{SP97}).
These terms were shown \cite{SPJR89} to be, in general, strongly suppressed
by the various averagings one has to perform when analyzing
the solar neutrino data in terms of the hypothesis
that solar neutrinos undergo matter-enhanced transitions in the Sun.
More specifically, it was found \cite{SPJR89}
that the oscillating terms in
$P_{\odot}(\nu_e \rightarrow \nu_{e};t_s, t_0)$
can be important only for the monochromatic $^7$Be-- and pep--neutrinos
and only for values of $\Delta m^2 \la 10^{-8}~{\rm eV^2}$. 
As we shall see, the current solar neutrino data suggest that
$\Delta m^2 \ga 10^{-7}~{\rm eV^2}$.

It should be emphasized that for $\Delta m^2 \ga 10^{-7}~{\rm eV^2}$ the
averaging over the region of solar neutrino production in the Sun
and the integration over the neutrino energy renders negligible all
interference terms which appear in the probability of $\nu_e$ survival due
to the $\nu_e\leftrightarrow\nu_{\mu (\tau)}$ oscillations in vacuum taking
place on the way of the neutrinos from the surface of the Sun to the
surface of the Earth. Thus, the probability that $\nu_e$ will
remain $\nu_e$ while it travels from the central part of the Sun to the
surface of the  Earth is effectively equal to 
the  probability of survival of the $\nu_e$ while it propagates
from the central part of the Sun to the surface of the Sun and is given
by the average probability
$\bar{P}_{\odot}(\nu_e \rightarrow \nu_{e};t_s, t_0)$
(determined by (62) and (63)).

The probability $\bar{P}_{\odot}(\nu_e \rightarrow \nu_{e};t_s, t_0)$
has several interesting properties. 
If the solar $\nu_e$ transitions are
adiabatic (i.e., $P' \cong 0$) and
$\cos2\theta_{m}(t_0) \cong -1$
(i.e., $N_e(t_0)/N_e^{res} \gg 1, \tan2\theta$, solar neutrinos are born
``above'' and ``far'' (in $N_e$) from the resonance region), one has
$$
\bar{P}(\nu_e \rightarrow \nu_e ; t_s, t_0) \cong
  \sin^2\theta,~~~\eqno(64)
$$
which is compatible with the qualitative result  (52a) derived
earlier. The solar $\nu_e$ undergo extreme nonadiabatic transitions in the
Sun ($4n_0 \ll 1$) if, e.g., $E/\Delta m^2$ is ``large'' 
(see  (60)). In this case  
again $\cos2\theta_{m}(t_0) \cong -1$ and, as it follows \cite{SP200} from  (63),
$P' \cong \cos^2\theta$. Correspondingly, the average probability 
takes the form:  
$$
\bar{P}(\nu_e \rightarrow \nu_e ; t_s, t_0) \cong
 1 - {1\over {2}}\sin^22\theta~,~~\eqno(65)
$$
which is the average 
two-neutrino vacuum oscillation probability.
Thus, if the solar neutrino transitions are extremely nonadiabatic,
the $\nu_e$ undergo oscillations in the Sun as in vacuum.
We get the same result, eq. (65),
if $N_e(t_0)(1 - \tan2\theta)^{-1} < N_e^{res}$, i.e.,
when $E/\Delta m^2$ is sufficiently small so that 
the resonance density exceeds 
the density in the point of neutrino
production. In this case \cite{KP88} the $\nu_e$ transitions
are adiabatic ($P' \cong 0$) and again
the $\nu_e \leftrightarrow \nu_{\mu (\tau)}$ oscillations 
take place in the Sun  as in vacuum:
$\cos2\theta_{m}(t_0) \cong \cos2\theta$ and
$\bar{P}(\nu_e \rightarrow \nu_e ; t_s, t_0) \cong 1 - {1\over {2}}\sin^22\theta$.

Let us note that the general aspects of the discussion and the 
results presented above are valid also in the case of 
solar neutrino transitions into sterile neutrino,
$\nu_e\rightarrow\nu_{s}$. 
In particular, the average probability
$\bar{P}_{\odot}(\nu_e \rightarrow \nu_{e};t_s, t_0)$
in this case 
is given effectively by (62) and (63) with \cite{LangP}
$N_e(t_0)$ replaced by $(N_e(t_0) - 
1/2N_n(t_0))$ in the expression for $\cos 2\theta_{m}(t_0)$,
$N_n(t_0)$ being the neutron number 
density of neutrons in the point of neutrino production in the Sun. 

The probability $\bar{P}(\nu_e \rightarrow \nu_e ; t_s, t_0)$ 
is shown as function of $E/\Delta m^2$ for three values of 
$\sin^22\theta = 0.8;~0.2;~5\times 10^{-3}$ in Figs. 7a - 7c.

\begin{figure}[p]
\begin{center}
\epsfig{file=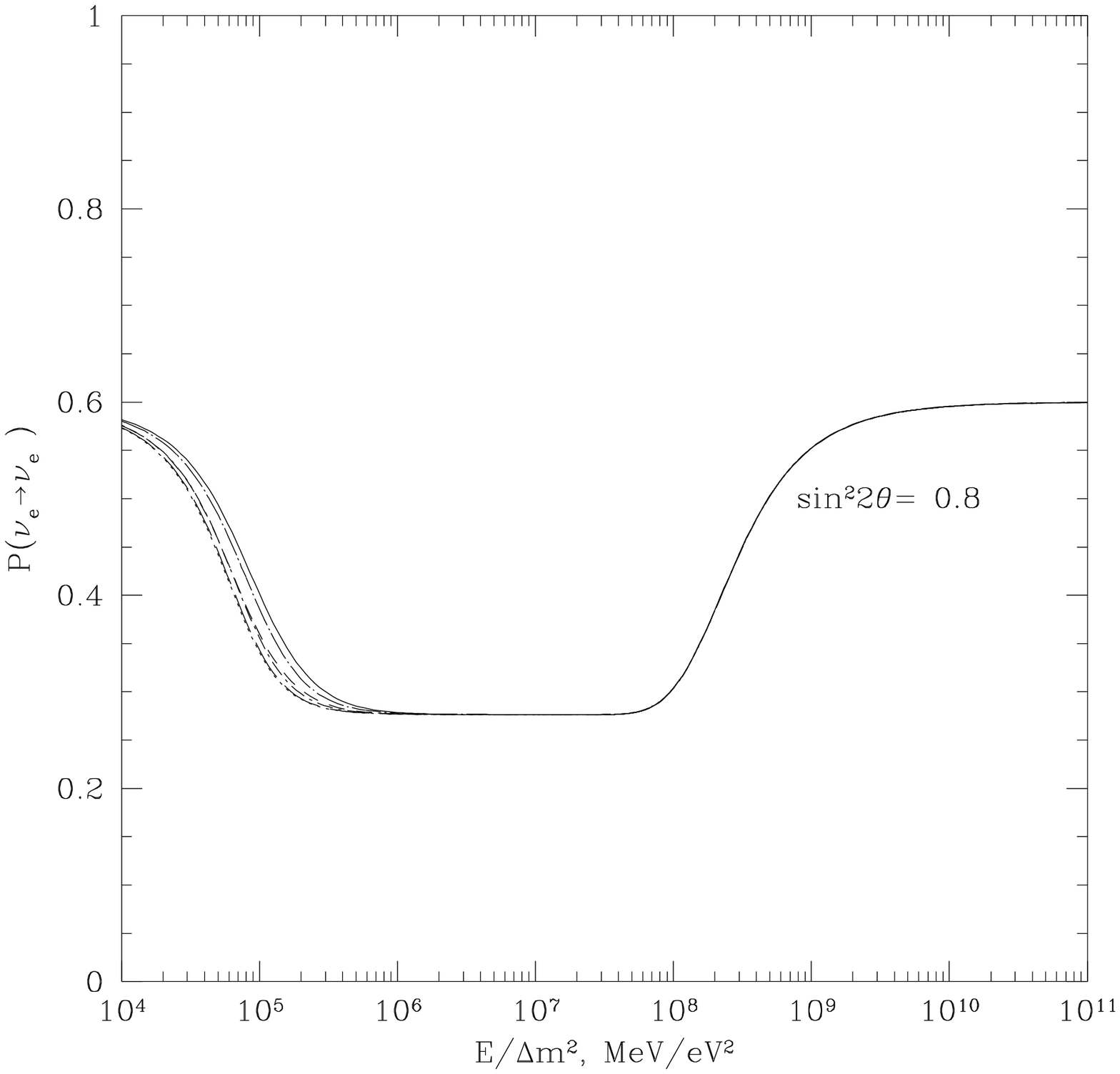,width=7cm}\\
\epsfig{file=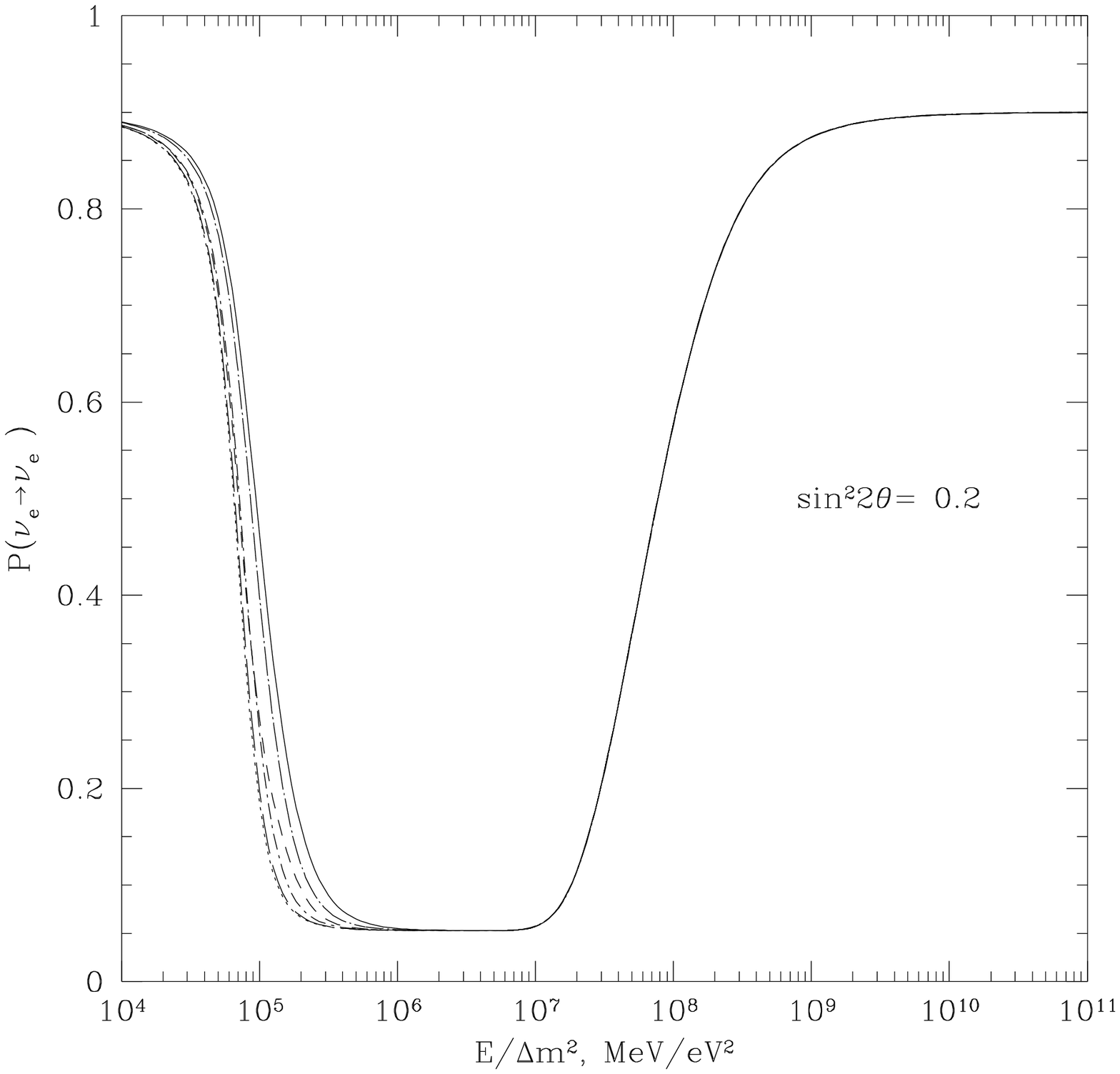,width=7cm}\\
\end{center}
\caption{(c)The solar $\nu_e$ survival probability [90]
$\bar{P}_{\odot}(\nu_e \rightarrow \nu_{e};t_s, t_0)$, Eq. (62),
averaged over the region of production the
pp (solid line), pep (long-dash-dotted line),
$^{13}$N (dashed line), $^{7}$Be (dash-dotted line),
$^{15}$O (long-dashed line)
and $^{8}$B (dotted line) neutrinos for $\sin^22\theta =
0.8~(a);~0.2~(b);~0.005~(c)$ as a function of $E/\Delta m^2$.
Figures a and b correspond to
$\nu_e \rightarrow \nu_{\mu(\tau)}$ transitions, while
figure c corresponds to $\nu_e \rightarrow \nu_{s}$
transitions.}
\end{figure}

\begin{figure*}[t]
\begin{center}
\epsfig{file=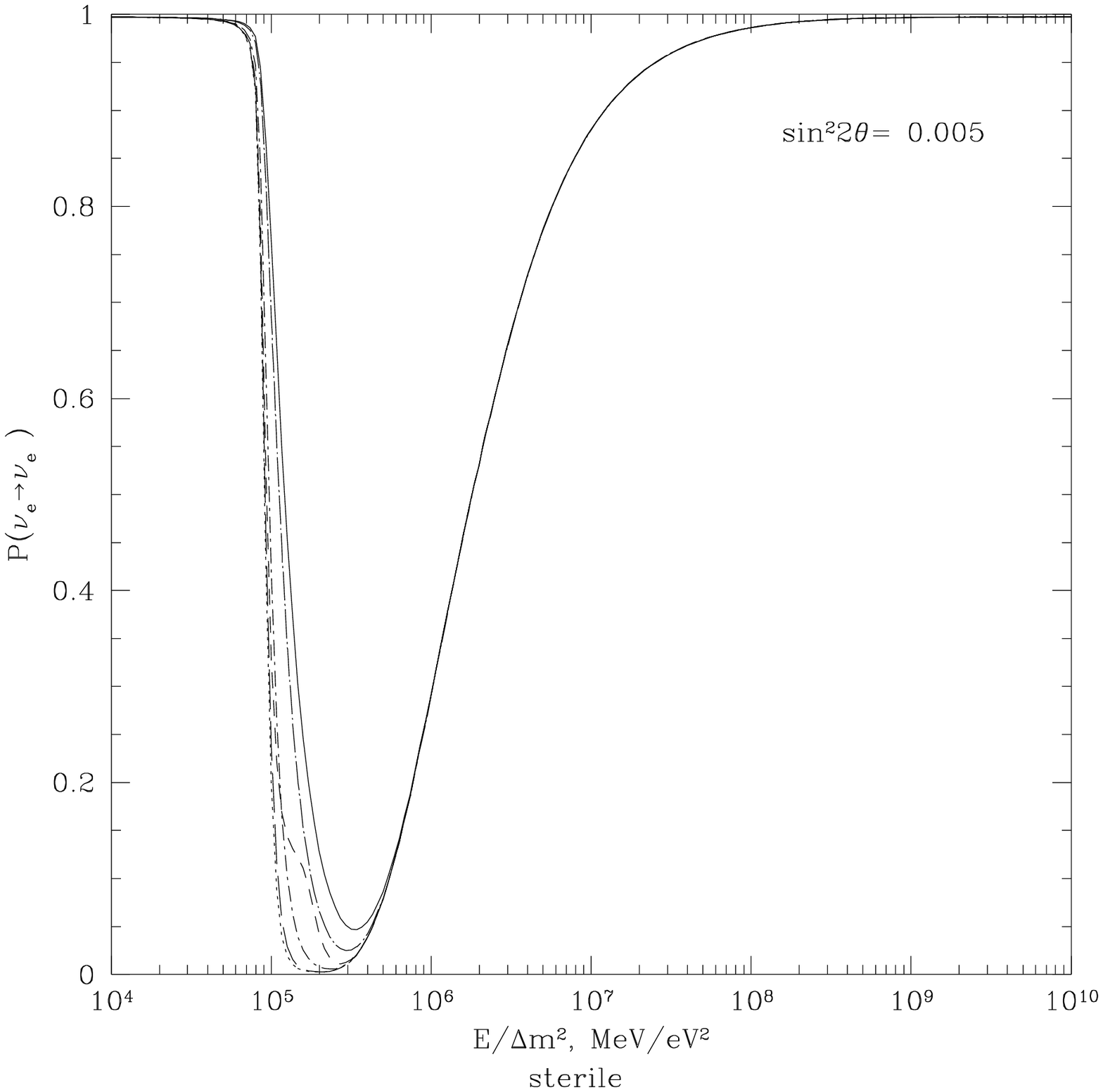,width=7cm}
\vspace{5mm}\\
{\bf Fig. 7} (c)
\end{center}
\end{figure*}

Further details concerning the analytic description of the matter--enhanced
transitions of the solar neutrinos in the Sun can be found in 
\cite{SP200}, \cite{SP191}--\cite{SP90}, \cite{LINEAR,SP97}.
Exact analytic results for the probability of 
various possible two-neutrino matter-enhanced 
transitions in a medium
($\nu_e\rightarrow\nu_{\mu (\tau)}$ or the inverse,
$\nu_e\rightarrow\nu_{\bar{\mu} (\bar{\tau})}$ or the inverse,
$\bar{\nu}_e\rightarrow\nu_{\mu (\tau)}$ or the inverse,
$\nu_{\mu}\rightarrow\nu_{s}$, etc.), which
are based solely on the general 
properties of the system of evolution
equations (32) (and do not make use of 
the explicit form of the functions
$\epsilon(t)$ and $\epsilon'(t))$ are given in \cite{SP97}.

Earlier studies (from 1993 -- 1994) of the possibility 
to explain the solar
neutrino problem in terms of the hypothesis of matter--enhanced
$\nu_e\rightarrow\nu_{\mu (\tau)}$ transitions of
solar neutrinos have shown \cite{KP4} that 
the data admits, in general, two types of
MSW solutions: a small mixing angle nonadiabatic solution for
$10^{-3} < \sin^22\theta \la 10^{-2}$, and a large mixing angle adiabatic
one for approximately $0.60 \la \sin^22\theta \la 0.95$, with the
allowed values of $\Delta m^2$ lying in
the interval $10^{-7}~{\rm eV^2} \la \Delta m^2 \la 10^{-4}~{\rm eV^2}$.
The terms ``nonadiabatic'' and ``adiabatic'' refer to the type of 
transitions the $^{8}$B neutrinos undergo in the corresponding cases.
It was also shown (see, e.g., \cite{KP2,KP5})
that in the case of $\nu_e\rightarrow\nu_{s}$ transitions
only a small mixing angle nonadiabatic solution, analogous to the
$\nu_e\rightarrow\nu_{\mu (\tau)}$
nonadiabatic solution, is allowed by the data.

Recently the MSW solutions of the solar neutrino problem have been
re-examined \cite{PK97,SPnu96,SPHeid97}~  
(exploiting the $\chi^2-$method) using the
data  (1), (2), and (4), the GALLEX result from 51 runs of
measurements, $\bar {\rm R}_{\rm GALLEX}({\rm Ge}) =
(69.7~\pm 6.7^{+~3.9}_{-~4.5}) \hskip 0.2cm {\rm SNU}$, and
the Super-Kamiokande result from 201.6 days of measurements ($\sim 3000$ events),
$\Phi^{{\rm SK}}_{{\rm B}} =
(2.65~~^{+0.09}_{-0.08}~~^{+0.14}_{-0.10}) \times 10^{6}~\nu_e/cm^{2}/sec~$.
The analysis was based on the predictions of the
solar model of \cite{BP95} with heavy element diffusion
for the electron and neutron number density distributions\footnote{All 
solar models compatible with the currently existing
observational constraints (helioseismological and other) predict 
practically the same electron and neutron 
number density distributions in the Sun.} and for the relevant 
pp, pep, $^{7}$Be, $^{8}$B
and CNO components of the solar neutrino flux.
The uncertainties in the predictions for the fluxes 
estimated in \cite{BP95} as well as 
the uncertainties
of the different solar neutrino detection reaction cross-sections
were taken into account.
The probability $\bar{P}_{\odot}(\nu_e \rightarrow \nu_{e};t_s, t_0)$
was calculated following the prescriptions given in \cite{KP88}.
The results obtained in the
cases of $\nu_e\rightarrow\nu_{\mu (\tau)}$ and
$\nu_e\rightarrow\nu_{s}$ transitions
are depicted in Figs. 8 and 9, respectively.

\begin{figure}[ht]
\begin{center}
\epsfig{file=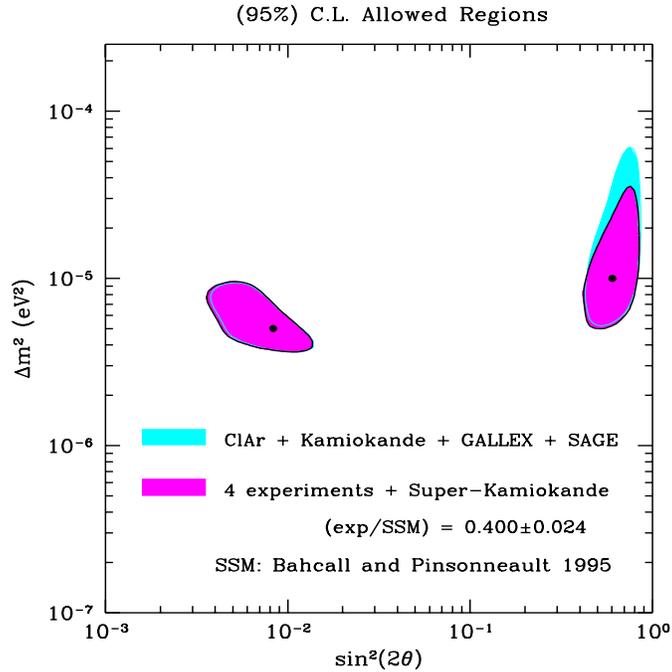,height=9cm,width=9cm}
\end{center}
\caption{Regions of values of the parameters 
$\Delta m^2$ and $\sin^22\theta$ (the black areas) for which 
the matter-enhanced $\nu_{e} \rightarrow \nu_{\mu (\tau)}$ 
transitions of solar $\nu_e$ allows to describe (at 95\% C.L.) the solar
neutrino data (from [72]
). For further details see the text.}
\end{figure}

\begin{figure}[ht]
\begin{center}
\epsfig{file=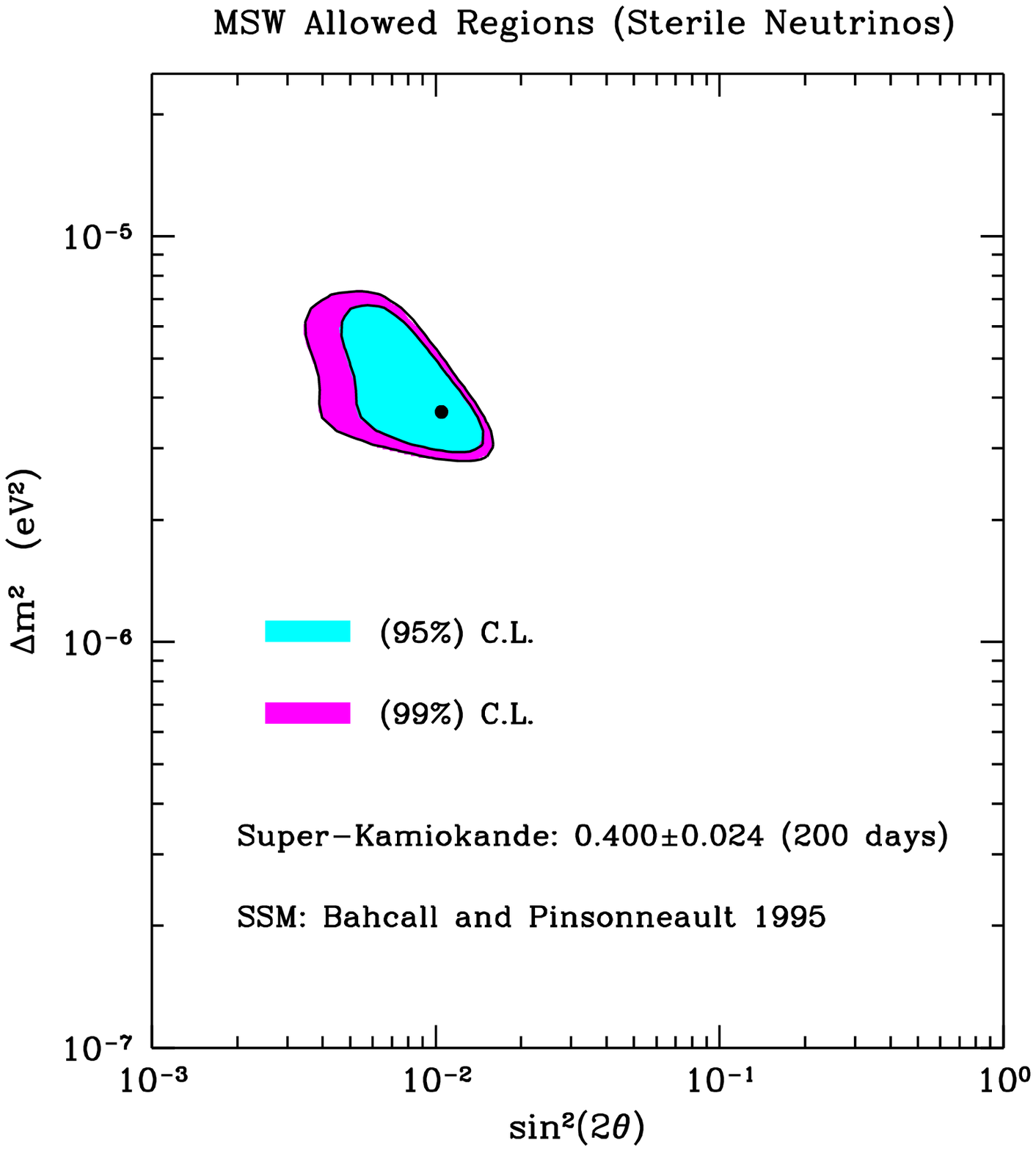,height=9cm,width=9cm}
\end{center}
\caption{
Allowed region of values of the parameters 
$\Delta m^2$ and $\sin^22\theta$ corresponding to the 
matter-enhanced $\nu_{e} \rightarrow \nu_{s}$
transition solution of the solar 
neutrino problem (from [72]
). For further details see the text. 
}
\end{figure}

The solid line contours in Fig. 8 denote 
regions allowed by the data
from the Homestake, Kamiokande, SAGE 
and GALLEX  experiments,
while the dark shaded
regions have been obtained by including the Super-Kamiokande
data in the analysis. Thus, the dark
shaded areas represent the regions allowed
by the mean event rate data from all experiments.
The solid line contours in Fig. 9 denote 
the region allowed by the data (at 95\% and at 99\% C.L.).

The current solar neutrino data are best described
assuming the solar neutrinos undergo
small mixing angle $\nu_{e} \rightarrow \nu_{\mu (\tau)}$
matter-enhanced
transitions \cite{PK97,SPnu96,SPHeid97}~
(for this  nonadiabatic solution one has $\chi^2_{min} = 0.9$ (3 d.f.)).
The quality of the fit of the data is somewhat worse
in the case of the
large mixing angle or adiabatic 
solution ($\chi^2_{min}$ is somewhat larger:
$\chi^2_{min} = 1.5$).
A similar quality of the fit of the data is provided also by
the hypothesis of transitions into a sterile neutrino,
$\nu_{e} \rightarrow \nu_{s}$, at small mixing angles ($\chi^2_{min} = 1.5$).
In contrast, the large mixing angle  $\nu_{e} \rightarrow \nu_{s}$
transition solution is practically excluded as a possible explanation
of the solar neutrino deficit \cite{PK97,SPnu96,SPHeid97}~ (it is
ruled out at 99.98\% C.L. ($\chi^2_{min} = 21$, 3 d.f.) by the data).

The values of the parameters $\Delta m^2$ and $\sin^22\theta$
for which one obtains (at 95\% C.L.) the
small mixing angle $\nu_e\rightarrow\nu_{\mu (\tau)}$ transition
solution of the solar neutrino problem lie in the region:
$$3.8\times10^{-6} {\rm eV}^2\la \Delta m^2 \la
10^{-5} {\rm eV}^2,~~\eqno(66a)$$
$$
3.5\times 10^{-3} \la \sin^22\theta \la 1.4\times 10^{-2}.~~~\eqno(66b)
$$
As Figs. 7 and 8 show, the small mixing angle
$\nu_e\rightarrow\nu_{s}$ solution region is very similar in shape
and magnitude to the region of the
$\nu_e\rightarrow\nu_{\mu (\tau)}$ solution, (66a) and (66b),
but is shifted with respect to the latter
by a factor of $\sim 1.3$ to smaller
values of $\Delta m^2$.

We have seen that there can be large uncertainties in the solar model
predictions for the total flux of $^{8}$B neutrinos and that the predictions
for the $^{7}$Be neutrino flux vary by $\sim$25\%. The question
of how stable are the MSW solutions of the solar neutrino problem
discussed above with respect to changes in the predictions for
the two fluxes ${\rm \Phi_{B}}$ and ${\rm \Phi_{Be}}$ naturally arises.
A rather comprehensive answer to this question
for the  $\nu_{e} \rightarrow
\nu_{\mu (\tau)}$ transition solution
was given in  \cite{KrSm}, and for the
solution with $\nu_e$ transitions into a
sterile neutrino, $\nu_{e} \rightarrow \nu_{s}$ - in
\cite{KP4}. These studies showed, in particular,
that the existence of the MSW solutions of the solar neutrino problem is
remarkably stable with respect to variations in the predictions for the $^{8}$B
and $^{7}$Be neutrino fluxes.

\subsection{A Detour: MSW Transitions of Solar Neutrinos in the Sun
and the Hydrogen Atom}

As we have indicated, the
two-neutrino matter-enhanced $\nu_e \rightarrow \nu_{\mu(\tau)}$
transitions of solar neutrinos at small mixing angles provide the best
description of the solar neutrino data.
In the present subsection we demonstrate \cite{SP97} that the second order
differential equation
for the probability amplitude
$A_{e}(t, t_0)$ of solar $\nu_e$ survival coincides in form
in the case of solar 
electron number density $N_e(t)$
changing exponentially along the neutrino path, Eq. (50),
with the Schr\"odinger equation for 
the radial part of the non-relativistic
wave function of the hydrogen atom, 
and we comment briefly on this interesting
coincidence.

Using the first equation in  (32) to express
$A_{\mu}(t,t_0)$ in terms of $A_{e}(t,t_0)$ and its time derivative, which gives
$A_{\mu}(t,t_0) = {1\over{\epsilon'}}~(\epsilon(t) + i{d\over{dt}})A_{e}(t,t_0)$,
and substituting $A_{\mu}(t,t_0)$ thus found in
the second equation in (32), we obtain a second
order differential equation for $A_{e}(t,t_0)$:
$$
\left\{~{d^2\over{dt^2}}
+ [\epsilon^2 + \epsilon'^2 - i\dot{\epsilon}]
\right\}A_{e}(t,t_0) = 0,~~~\eqno(67)
$$
where $\dot{\epsilon} = {d\over{dt}}\epsilon$ and
$\epsilon(t)$ and $\epsilon'$ are given by  (33).
Introducing the dimensionless variable
$$Z = ir_0\sqrt{2}G_{F}N_e(t_0)e^{- {{t - t_0}\over{r_0}}},~~~
 Z_0 = Z(t = t_0),~~~\eqno(68)$$
and making the substitution
$$A_{e}(t,t_0) \equiv A(\nu_{e}\rightarrow \nu_{e}) =
(Z/Z_0)^{c-a}~e^{ -(Z - Z_0) + i\int_{t_0}^{t}\epsilon (t')dt'}~A'_{e}(t,t_0),
~~~\eqno(69)$$
we find that the amplitude $A'_{e}(t,t_0)$ satisfies
\cite{SP200,EXP,SP88osc} the
confluent hypergeometric equation \cite{BE}:
$$
\left\{~Z{d^2\over{dZ^2}} + (c - Z)~{d\over{dZ}}
- a \right\}A'_{e}(t,t_0) = 0,~~\eqno(70)
$$
where \cite{SP88osc}
$$
a = 1 + ir_0~{{\Delta m^2}\over{2E}}\sin^2\theta,~~~~
c = 1 + ir_0~{{\Delta m^2}\over{2E}}.~~~\eqno(71)
$$

Equation (70) coincides in form with the Schr\"odinger
(energy eigenvalue) equation obeyed by the radial part,
$\psi_{kl}(r)$, of the non-relativistic wave function of the
hydrogen atom \cite{CT}, $\Psi(\stackrel{\rightarrow}{r}) =
{1\over{r}}\psi_{kl}(r)Y_{lm}(\theta',\phi')$,
where $r$, $\theta'$ and $\phi'$ are the spherical coordinates of the electron
in the proton's rest frame,
$l$ and $m$ are the orbital momentum quantum numbers ($m = -l,...,l$),
$k$ is the quantum number labeling (together with $l$) the 
electron energy,\footnote{The principal quantum number 
is equal to $(k + l)$ \cite{CT}.}
$E_{kl}$ ($E_{kl} < 0$),
and $Y_{lm}(\theta',\phi')$ are the spherical harmonics. To be more precise,
the function $\psi'_{kl}(Z) = Z^{-c/2}~e^{Z/2}~ \psi_{kl}(r)$ 
satisfies equation (70), 
where the variable $Z$ and the parameters $a$ and $c$
are in this case related to the physical quantities 
characterizing the hydrogen atom:
$$
Z = 2~{r\over{a_0}}\sqrt{- E_{kl}/E_{I}},~~
a \equiv a_{kl} = l + 1 - \sqrt{- E_{I}/E_{kl}},~~
c \equiv c_{l} = 2(l + 1).~~\eqno(72)
$$
Here $a_0 = \hbar/(m_ee^2)$ is the Bohr radius and
$E_{I} = m_ee^4/(2\hbar^2) \cong 13.6~eV$ is the
ionization energy of the hydrogen atom. It is remarkable that
the behaviour of such different physical systems as solar neutrinos
undergoing matter-enhanced transitions in the Sun and the non-relativistic
hydrogen atom are governed by one and the same differential equation.

The properties of the linearly independent
solutions of equation (70), i.e., of the confluent hypergeometric
functions, $\Phi(a,c;Z)$, as well as their asymptotic series
expansions, are well-known
\cite{BE}. Any solution of (70) can be expressed as a linear
combination of two linearly independent solutions of (70),
$\Phi(a,c;Z)$ and $Z^{1 - c}~\Phi(a-c+1,2-c;Z)$, which are
distinguished from other sets of linearly independent confluent
hypergeometric functions by
their behaviour when $Z \rightarrow 0$:
$\Phi(a',c';Z = 0) = 1$, $a',c' \neq 0, -1, -2,...$,
$a'$ and $c'$ being arbitrary parameters. Explicit expressions for the
probability amplitudes $A(\nu_{e}\rightarrow \nu_{e})$ and
$A(\nu_{e}\rightarrow \nu_{\mu (\tau)})$ in terms of the functions
$\Phi(a,c;Z)$ and $\Phi(a-c+1,2-c;Z)$ were derived in \cite{SP88osc,AASP92}.
In the case of MSW transitions of solar neutrinos
($N_e(t_s) = 0)$ these expressions 
have an especially simple form: they are given by
the corresponding vacuum oscillation probability 
amplitudes ``distorted'' by the values of the
functions $\Phi(a',c';Z)$ in the initial 
point of the neutrino trajectory,
$$
A(\nu_{e}\rightarrow \nu_{\mu (\tau)})
= {1\over{2}}~\sin2\theta~\left \{\Phi(a-c,2-c;Z_0) - e^{i(t - t_0)
{{\Delta m^2}\over{2E}}}~\Phi(a-1,c;Z_0)~\right \},~~~\eqno(73)
$$
etc., where $Z_0$, $a$ and $c$ are defined in (68) and (71).
In the limit $|Z_0| \rightarrow 0$, which corresponds
to zero electron number density,
expression (73) reduces (up to an irrelevant common phase factor) 
to the one for oscillations in vacuum, Eq. (23).

It is well-known that the requirement of a correct asymptotic behaviour
of the wave function $\psi_{kl}(r)$ at large $r$ leads to the quantization
condition for the energy of the electron, $E_{kl}$, 
in the hydrogen atom \cite{CT}~:
$E_{kl} = - E_{I}/(k + l)^2$, $(k + l) = 1,2,...$ ($l = 0,1,2,...,(k+l)-1$).
Technically, the condition is derived by 
using the asymptotic series expansion
of the confluent hypergeometric functions 
in inverse powers of the argument $Z$ \cite{BE}
(one has $Z \rightarrow \infty$ 
when $r \rightarrow \infty$, see  (72)).
The same asymptotic series
expansion in the case of the solutions describing the
MSW transitions of solar neutrinos in the Sun (we have $|Z_0| \ga 520$
in this case \cite{SP88osc}) permitted to derive i) the simple
expression for the relevant ``jump'' probability \cite{SP200} $P'$, Eq. (63), 
and ii) explicit expressions for
the oscillating terms in the solar $\nu_e$ survival probability
\cite{SP88osc}. Expression (63) is a basic ingredient of the most
precise simple analytic description of the two-neutrino
matter-enhanced transitions of solar neutrinos in the Sun, 
available at present \cite{KP88}.

\section{The Solar Neutrino Problem: Outlook}

After being with us for $\sim$25 years the solar neutrino problem still
remains unsolved. With the accumulation of the quantitatively new data provided
by the Ga--Ge experiments the problem acquired a novel aspect: the constraints
on the $^{7}$Be neutrino flux following from the data imply a significantly
smaller value of $\Phi_{Be}$ than is predicted by the solar models. The
data of both Davis et al. and Kamiokande experiments have to be incorrect in
order for the indicated conclusion to be not valid. The vacuum oscillations and
MSW transitions of the solar neutrinos continue to be viable and very
attractive solutions of the problem.

The start of the Super-Kamiokande experiment on April 1, 1996, and
the presentation of the first preliminary data from this experiment
at the ``Neutrino '96'' International Conference in June of the same year
\cite{SK}, marked the beginning of a new era in the
experimental studies of solar neutrinos. This is the
era of high statistics experiments with real time event detection
and capabilities to perform high precision
spectrum, seasonal variation \cite{NMP2,KP2}, day-night asymmetry
(see, e.g., \cite{DN1,DN2} and the articles quoted therein), etc.,
measurements. Such capabilities are of crucial importance,
in particular, for understanding
the true cause of the solar neutrino deficit.

The preceding period 1967 - 1996 of solar neutrino measurements,
which began when the epic Homestake (Cl-Ar)
experiment started to collect data \cite{Davis68,Davis},
is marked by several remarkable achievements
which, given their scale and the time and the efforts
they took, make this period rather an epoch.
For the first time neutrinos emitted by
the Sun have been observed. The thermo-nuclear reaction theory of
solar energy generation was confirmed by the detection by GALLEX and SAGE
experiments of the lower energy solar neutrinos
produced in the corresponding fusion nuclear reactions.
More generally, this result confirms a fundamental aspect of the
theory of stellar evolution regarding the role played by
the nuclear fusion reactions.
Finally, the solar neutrino data gathered in 
the indicated period
provided, when compared with the 
predictions of the solar models,
indirect evidences for an ``unconventional'' behaviour
(e.g., vacuum oscillations, and/or matter-enhanced transitions, etc.)
of the solar neutrinos on their way to the Earth.
This in turn is the strongest indication we presently
have for the existence of new physics beyond that predicted
by the standard theory of electroweak and strong interactions.

The Super-Kamiokande is the first
operating of a group of new generation detectors,
SNO \cite{SNO}, BOREXINO \cite{BOREXINO}, 
ICARUS \cite{ICARUS}, HELLAZ \cite{HELLAZ}, etc.,
which will allow one to perform more
detailed and accurate studies of the solar
neutrino flux reaching the Earth.
As is well known, Super-Kamiokande, SNO and ICARUS experiments
will study the $^{8}$B component of the solar neutrino flux at energies
of solar neutrinos $E \ga (5 - 6)~{\rm MeV}$;
the BOREXINO detector is designed to provide information
about the 0.862 MeV $^{7}$Be component of the flux:
approximately 90\% of
the signal produced by the solar 
neutrinos in the BOREXINO
detector ($\sim$50 events/day according to 
the reference model \cite{BP95}) is
predicted to be due to the $^7$Be--neutrinos. The HELLAZ
apparatus is envisaged to measure the total flux and the spectrum
of the pp neutrinos\,\footnote{The HELLAZ detector can be utilized for
studies of the $^{7}$Be neutrino flux as well.} in the energy interval 
$E \cong (0.22 - 0.41)~{\rm MeV}$.

The SNO experiment
is expected to begin to take data in 1998.
The construction of the BOREXINO detector is under way and
is planned to be completed by the end of 1998.
A prototype of the ICARUS apparatus has been successfully
tested and the construction of the first 600 ton module has started.
The feasibility studies for the HELLAZ detector have been
intensified with the building of a small
prototype at College de France
\cite{HELLAZ}. Our aspirations to find the cause of the solar neutrino
deficit established by the results of the spectacular
solar neutrino experiments of the first generation
\cite{Davis68,Davis}, \cite{Kam}--\cite{GALLEX},
and confirmed by the first results 
from the Super-Kamiokande detector, 
and to get additional independent
information about the physical conditions in the central
part of the Sun, are presently associated with
the more precise and diverse data the second generation detectors
are expected to provide. All these are planned to be
high statistics (typically $\sim$3000 solar neutrino 
events/year, Super-Kamiokande is expected to collect
$\sim$10000 events/year), i.e., high precision, 
experiments  with real time event detection. 

In SNO experiment the $^{8}$B neutrinos will be detected via the 
charged current and the neutral current reactions on deuterium: 
$\nu_{e} + {\rm D \rightarrow e^{-} + p + p}$, and 
$\nu + {\rm D \rightarrow \nu + p + n}$; the measurement of the kinetic 
energy of the electron in the first reaction will permit to 
search for possible deformations of the 
spectrum of $^{8}$B neutrinos at $E \geq 6.44~$MeV, 
predicted to exist (see, e.g., the first article quoted in 
\cite{KP4} as well as \cite{KP2,KP5}) if solar neutrinos take part in
oscillations in vacuum on the way to the Earth 
and/or undergo matter-enhanced transitions in the Sun.
High precision searches for spectrum deformations 
will be performed  also in the Super- 
Kamiokande experiment in which the energy of the recoil electron from the 
$\nu - {\rm e}^{-}$ elastic scattering reaction will be measured with a 
high  accuracy. 
 
The high statistics these experiments will
accumulate, the measurement of the spectra
of final state electrons with the
SNO and Super Kamiokande detectors, and of 
the ratio of the charged current and
the neutral current reaction rates with 
the SNO detector, will make it 
possible to perform various critical tests 
(see, e.g., \cite{KP2,KP5,DN1,DN2})  
of the vacuum oscillation and the 
MSW, as well as of the other possible 
neutrino physics solutions 
\cite{RSFP1}--\cite{FCNC2,DEC1}  of 
the solar neutrino problem. We may be at the dawn of a major breakthrough in 
the studies of solar neutrinos. It is not excluded, however, 
that the data from the BOREXINO and
HELLAZ detectors may be required 
to get an unambiguous answer concerning the
cause of the solar neutrino problem \cite{KrSm,KP3,KP5}.

\subsubsection{Acknowledgements.}
It is a pleasure to thank
the organizers of the 36. Internationale Universit\"atswochen
f\"ur Kern- und Teilchenphysik 1997 in Schladming
for the enjoyable atmosphere created at the School.

\end{document}